\documentclass[utf8]{FrontiersinHarvard} 

\DeclareUnicodeCharacter{03B1}{ }
\usepackage[]{url,lineno,microtype}
\usepackage[backref=page]{hyperref}
\usepackage[onehalfspacing]{setspace}
\usepackage{natbib}
\usepackage{comment}
\newcommand{\ca}{\mbox{Ca\,{\sc ii}~K\,}}

\def\keyFont{\fontsize{8}{11}\helveticabold }
\def\firstAuthorLast{Ermolli  {et~al.}} 
\def\Authors{Ilaria~Ermolli\,$^{1,*}$, Fabrizio~Giorgi\,$^{1}$,  Theodosios~Chatzistergos\,$^{2}$}
\def\Address{$^{1}$INAF Osservatorio Astronomico di Roma, via Frascati 33, 00078 Monte Porzio Catone, Italy\\
$^{2}$Max Planck Institute for Solar System Research, Justus-von-Liebig-Weg 3,	37077 G\"{o}ttingen, Germany}
\def\corrAuthor{Ilaria Ermolli}

\def\corrEmail{ilaria.ermolli@inaf.it}

\begin{document}
\onecolumn      
\sloppy
\firstpage{1}

\title[Rome/PSPT]{Rome Precision Solar Photometric Telescope: precision solar full-disk photometry during solar cycles 23-25}
\author[\firstAuthorLast ]{\Authors} 
\address{} 
\correspondance{} 

\extraAuth{}

\maketitle
\begin{abstract}
\section{The Rome Precision Solar Photometric Telescope (Rome/PSPT) is a ground-based telescope engaged in precision solar photometry. It has a 27-year database of 
full-disk  images of the photosphere and chromosphere beginning in 1996 and continuing to 2022. The solar images have been obtained daily, weather permitting, with approximately 2 arcsec/pixel scale in \ca line at 393.3 nm, G-band at 430.6 nm, and continuum in the blue and red parts of the spectrum at 409.4 nm and 607.2 nm, respectively. Regular observations were also performed at the green continuum at 535.7 nm for a period of about 18 months.   Since the first-light, Rome/PSPT operations have been directed at understanding the source of short-and long-term 
solar irradiance changes, spanning from one minute to several months, and from one  year to a few solar cycles, respectively. However, Rome/PSPT data have also served to study a variety of other topics, including the photometric properties of solar disk features   
and of the  supergranulation manifested by  the 
chromospheric network. Moreover, they have been unique in allowing to connect series of historical and modern full-disk solar observations, especially  the   
\ca line data. 
Here, we provide an overview of the Rome/PSPT telescope and of the solar monitoring carried out with it  from its first light to the present, across solar cycles 23--25. We also briefly describe the main results achieved with Rome/PSPT data, and give an overview of new results being derived with the whole time series of observations covering the period 1996--2022.}
\tiny
 \keyFont{ \section{Keywords:} Sun: photosphere, Sun: chromosphere, Sun: sunspots, Sun: faculae, plages, Sun: activity, Astronomical instrumentation, methods and techniques} 
\end{abstract}

\section{Introduction}
Solar activity has been 
a key astrophysical research topic for many years. However, it has received even wider recognition since it became clear that the solar activity affects the Sun's radiative emission, with impact on the heliosphere and terrestrial conditions \citep[e.g.][]{haigh1996,gray2010,solanki2013,shindell2020,ipcc2021}.  
Indeed, the Sun's radiative emission represents the dominant energy source heating the Earth's atmosphere and climate \citep{kren2017}. For long time, this emission has been considered stable and even referred to as \textit{solar constant}  \citep{frohlich2010}.  However, measurements collected since 1978    
by a series of space-borne instruments have revealed that both the Total Solar Irradiance (TSI)  
and Spectral Solar Irradiance (SSI) 
vary at all timescales from minutes to decades, and  probably also on longer-term scales. We recall that the TSI is the solar radiative energy flux per unit area integrated over the entire spectrum,  measured at normal incidence at the top of the Earth’s
atmosphere and at a mean Sun-Earth distance of one astronomical unit, given in units of W m$^{-2}$, while the SSI is the analogous quantity but spectrally resolved,    
given in units of W m$^{-2}$nm$^{-1}$.  
All existing measurements show a clear 
TSI change by $\approx$ \ 0.1\% in phase with the 11-year solar activity cycle  and TSI fluctuations by up to 0.2–0.3\% on timescales shorter than a few days 
\citep[e.g.][]{frohlich2013,kopp2016,kopp2021,montillet2022};  
SSI changes in the UV in phase with the solar cycle are also well established and assessed to reach several tens of percent for the radiation below 200 nm 
\citep[e.g.][]{deland2008,woods2018,marchenko2019,woods2022}, 
 while there is still an uncertainty in both the phase and variability of the SSI changes in the visible and infrared ranges \citep{ermolli2013,coddington2019,dudok2022}. Still debated is also the TSI trend on the timescales longer than the solar cycle \citep[e.g.][]{kopp2021,schmutz2021}.  

In spite of limited data, soon after the beginning of the regular monitoring   of the solar irradiance  
the measured variations of TSI and SSI 
 were found to be clearly  associated 
with the brightness features generated by magnetic field emerged into the Sun's atmosphere \citep[e.g.][]{foukal1979,willson1981,foukal1986} and expression of the solar activity. In particular, the passages above the solar disk of magnetic features in the form of dark sunspot groups were seen to be associated with large dips recorded in the measured irradiance, while the passages of magnetic features with bright facular regions were found to be connected to irradiance increases. Early attempts to explain the 
measured changes in irradiance using observational data of the above solar disk features, and of other structures observed on the solar disk as dark pores and bright network, met however with only partial success \citep{chapman1987}.  
Indeed, solar observations and irradiance models available at that time were not  accurate enough to  test the measured variations against the various potential mechanisms of solar variability. These include e.g. 
Sun's structural changes in the convection zone and superadiabatic layer also counting the photosphere, in addition to surface magnetism  \citep[e.g.][]{newkirk1983,hudson1988,kuhn1988,kuhn2000,kuhn2004}. Improvements to observational data were needed to advance knowledge of the solar irradiance changes, and    
spatially resolved precision solar photometry was identified as a key input for new research in the field.

In the above framework, and with the increasing evidences of the Earth's global warming \citep[][]{ipcc1990}, in 1990 the U.S. National Science Foundation (NSF) \textit{Radiative Inputs from the Sun to the Earth} (RISE) project devoted to understanding the mechanisms of solar luminosity variations published  a research plan recommending  
the 
``design, construction and deployment of at least two solar photometric telescopes, optimized for precision photometric imaging of sunspots, faculae, and other photospheric features''. The NSF RISE project  later  founded a program at the U.S. National Solar Observatory at Sacramento Peak (NSO/SP) to develop a network of two or three specially designed telescopes to obtain full-disk solar photometric data \citep{kuhn1993}. In 1994 NSO/SP began development of a prototype of such telescopes, named \textit{Precision Solar Photometric Telescope} (PSPT). 

Since the main scientific interests connected with the use of the PSPT data concerned the understanding of the origin of the variations of solar irradiance, the main request of the telescope design was the accuracy of the differential photometry of the Sun's surface (both by pixel and observation) such as to allow a direct comparison with the variations of solar irradiance measured from space, of the order of 0.1\% of solar irradiance on short- and long- time scales. Thus the PSPT telescope was designed around a 15-cm objective and a large format CCD recording device for full-disk solar surface photometry with 0.1\% per pixel accuracy at seeing-limited spatial resolution \citep{coulter1994}.

In March 1995 the Osservatorio Astronomico di Roma (OAR), in collaboration with the Physics Department of the Universit\`a degli Studi di Roma ``Tor Vergata'', started a collaboration with the  RISE/PSPT project at 
NSO/SP  
aimed at realizing the first PSPT. 

After the installation of the first PSPT at the Rome Monte Mario site 
of the OAR 
(Rome, Italy; hereafter referred to as Rome/PSPT prototype\footnote{Latitude 41$^{\circ}$55'N, longitude 12$^{\circ}$27'E, elevation about 130 m. 
}), in February 1996 \citep{ermolli1998}, two other PSPT instruments were realized by NSO/SP and then installed, respectively in summer 1998 at the Mauna Loa Solar Observatory  (Hawaii, USA; hereafter referred to as MLSO/PSPT) under the operation of the High Altitude Observatory (HAO) of the U.S. National Center for Atmospheric Research  (NCAR), see, e.g., \citet{white2000}, and in year 2000 at the NSO/SP Observatory (New Mexico, USA;  hereafter referred to as NSOSP/PSPT), which also managed that telescope, see, e.g., \citet{kuhn1999}. The three telescopes have however had different evolution. Following the commissioning phase, the Rome/PSPT started the regular monitoring of the solar atmosphere in July 1996 and continued it until March 2022, but during a few periods of required maintenance of the telescope. 
The MLSO/PSPT, on the other hand, was operated from 1998 to 2015, 
and the NSOSP/PSPT solely acted as a spare instrument, never entering scientific operation.

Operation of the Rome/PSPT prototype and analysis of the data obtained with it in the first years of activity contributed to the final design of the PSPT telescopes and optimization of the procedures to reach the RISE/PSPT project goals, especially for the accurate photometric calibration of the data  acquired and quantification of their quality. This led, in June 1997, the Rome/PSPT prototype to be replaced by a quasi-definitive version of the instrument that was further upgraded in September 2001. At that time, the telescope was partly modified, so as to make it akin with the telescopes installed at the other sites. Besides,  following on 
changes in the organization of OAR, the Rome/PSPT was moved to the Monte Porzio Catone site\footnote{Latitude 41$^{\circ}$48'N, longitude 12$^{\circ}$42'E, elevation about 400 m.} 
of the OAR in the \textit{Castelli Romani} region south of Rome. 
Similarly, the MLSO/PSPT was upgraded over time, especially in the first years of its activity, and reached final set-up after multiple changes only seven years after the  first-light, in 2005.

Since 2001, the Rome/PSPT has differed in minor hardware characteristics to the MLSO/PSPT. Nevertheless, the existing differences have led to operational strategies implemented quite differently at the two telescopes. For example, the data acquisition with the Rome/PSPT has always been  operated by the observer and adapted to weather and observing conditions, while MLSO/PSPT data were acquired automatically within predefined time windows and with fixed exposure times for most of the MLSO/PSPT operation. 
In general, there was no coordination between the two telescopes, but to allow the continuation of the observations at the two sites. 
However, the data acquired with both telescopes have often been jointly investigated by the two groups managing the instruments, and analysed with common methods.

The partnership of OAR with NSO/SP for the realization of the Rome/PSPT, and further collaboration of HAO with NSO/SP for the installation of the MLSO/PSPT, allowed to realize the goal of the RISE/PSPT 
project of deploying at least two  
telescopes operating at widely separated sites for regular precision photometric imaging of the Sun's atmosphere. It is worth mentioning that the OAR participation in the RISE/PSPT project was supported by the Italian Ministry of Environment as a step towards a better  assessment of the role of Sun's  variability to the Earth's climate change. OAR, which became part of the Italian Istituto Nazionale di Astrofisica (INAF) in year 2000, has operated the Rome/PSPT in light of the NSF RISE program and in the framework of other national and international projects aimed at understanding the processes responsible for solar irradiance variations and solar activity. This was made possible thanks to the support received over years by the Italian Ministry of Education and Research, Regione Lazio, and  European Commission under  projects financed in the FP7 and H2020 programs.

In accordance with the goal of the NSF RISE program, since the telescope's first-light,
Rome/PSPT observations have been directed at understanding the source of  
solar irradiance changes through acquisition of daily images of the photosphere and chromosphere at various spectral bands. 
However, Rome/PSPT  data have also been acquired for the purposes of other research topics than the study of solar variability. For instance, there have been high-cadence observing runs, also coordinated with other telescopes, as e.g. during the campaigns performed in the frame of the Whole Heliosphere and Planetary Interactions (WHPI) initiative \citep{detoma2021}. Besides, the Rome/PSPT has monitored astronomical events as e.g. partial solar eclipses and transits of Venus and Mercury above the solar disk. Finally, over the years the Rome/PSPT 
has also served intense activity of dissemination of solar physics and solar-terrestrial relations at schools and general public.

Recently, a major hardware failure occurred in the control system of the Rome/PSPT that, at present, prevents collection of new observations. The ongoing recovery of the failed components will lead to an upgraded telescope with control  technologies for automated and remote operations and novel components, while maintaining the main features of the facility.

In this paper, we give an overview of the Rome/PSPT telescope and solar monitoring performed with it since 1996 and 
across solar cycles 23--25, as well as of the data acquired, their quality, and relevant results achieved from their analysis.  The discussion of the latter, which extend beyond the scope of the present paper, can be found in the original studies presented in the following.  This paper is structured as follows. After this introduction to the Rome/PSPT project, 
in Section 2 we describe the telescope and data acquisition. Sections 3 and 4 provide information on photometric calibration of the observations and their quality, respectively.  In Section 5 we present the data archive. Section 6 describes relevant  results achieved from  analysis of Rome/PSPT observations and new findings from the analysis of the whole time series of data. 
Conclusions are presented in Section 7.

\section{Telescope and Data acquisition}

\begin{figure*}
\begin{center}
\includegraphics[width=0.9\linewidth]{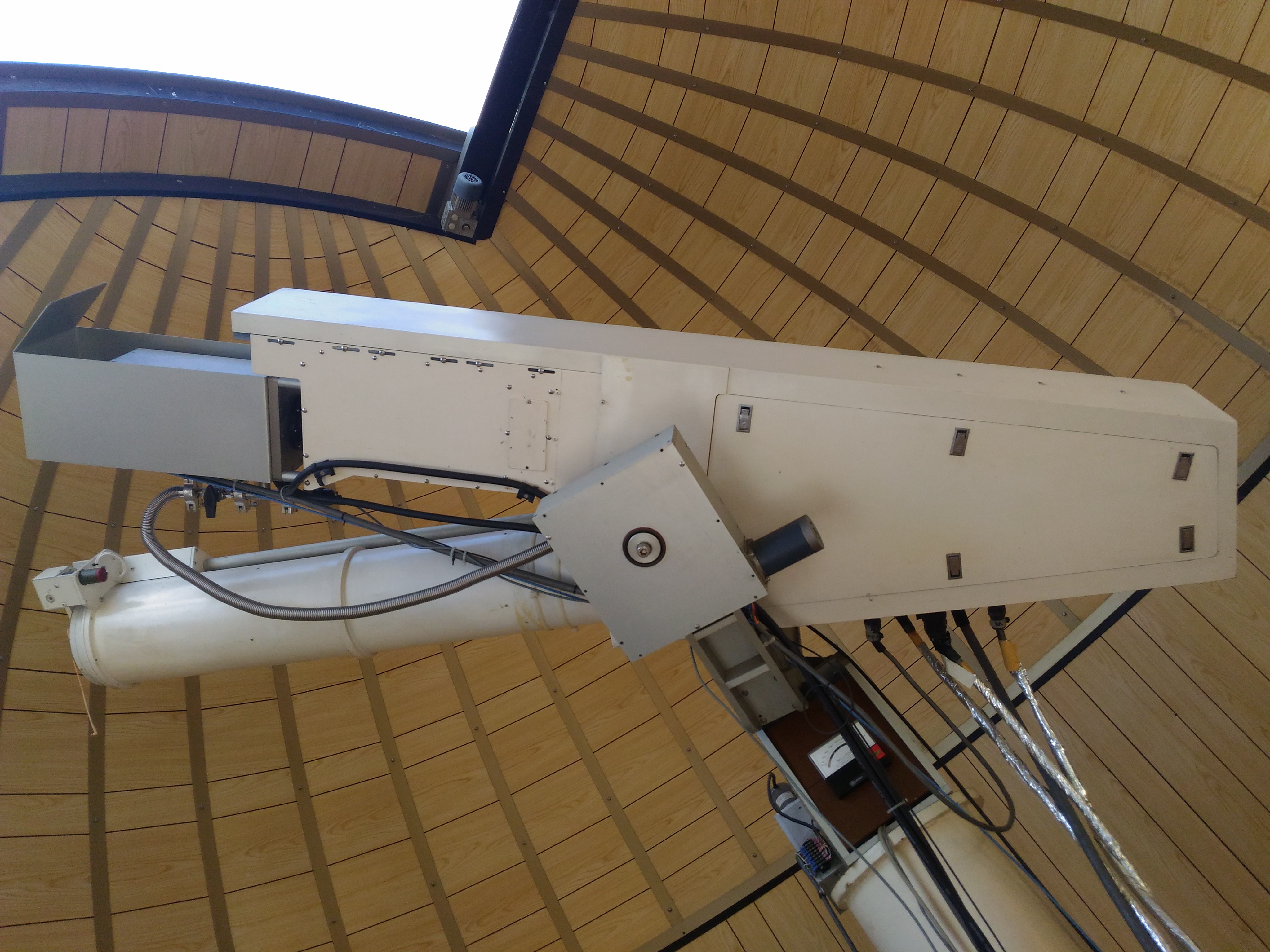}
\end{center}
\caption{The Rome/PSPT telescope in the host-dome at the INAF Osservatorio Astronomico di Roma in Monte Porzio Catone (Rome, Italy). The telescope has an equatorial mount and body comprising a lower tube that includes the objective lens and its dust cover, and upper part with the secondary optics and appended recording device. The guiding system is located in the back part of the telescope.
}\label{f1}
\end{figure*}

The Rome/PSPT is a 15-cm, low-scattered-light, refracting telescope designed for synoptic solar observations characterized by 0.1\% pixel-to-pixel relative photometric precision  at seeing-limited spatial resolution \citep{coulter1994}.   
It has an equatorial mount with a  computer-controlled pointing and  automation of all main components, and a folded and compact optical path. The pointing system operates on the principle of compensation based on electro-optical-mechanical feedback of the guiding elements described in the following.

Figure \ref{f1} shows the Rome/PSPT  during operations at the INAF OAR Monte Porzio Catone site. It consists of five major blocks: 1) objective lens, 2)  guiding system,  
3) secondary optics, 
4) recording device, and 5) control system.

The objective lens, with a 15 cm aperture and focal length of about 230 cm, comprises an achromatic doublet with an anti-infrared high-reflectance coating. The objective gives a 22 mm image of the Sun in the focal plane and detector. The anti-infrared coating reflects the   radiation above 700 nm, in order to reduce the heating of mechanical and optical components of the telescope. 

The guiding system consists of a
beam-splitter,  a Nikon 35 mm objective lens, a 1 cm quadrant-cell for telescope pointing, and a tip-tilt mirror of 5 cm diameter. The latter element is used to compensate the global  flickering motion of the solar image due to atmospheric seeing. The Nikon objective forms an image of the Sun on the quadrant-cell. This image, which is  slightly larger than 7 mm,  
is occulted in the central region in order to increase the sensitivity of the pointing system to small image shifts. Indeed, the image stabilization achieved with the guiding system allows summing frames in the recording device during observations, and thus increasing the signal to noise ratio of data, without a significant loss of spatial resolution. The quadrant-cell is placed close to the tip-tilt mirror in the back part of the telescope. According to the data acquired during usual observations, under clear sky conditions the guiding system works with an accuracy better than 1 arcsec.

The secondary optics system, which receives the  radiation mirrored by the tip-tilt element and transmitted by the beam splitter,  comprises an Uniblitz curtain shutter, a filter wheel carrying five interference filters, two doublet lenses that  form the collimated and focused beams at the entrance and exit of the interference filters, respectively, and a rotating shutter to guarantee the  
uniform illumination of the recording device.  
All the interference filters, which have a diameter of 50 mm, were manufactured  by Barr Associates Inc., with an excellent stability in the long-term properties of their transmission and a low sensitivity to temperature changes (band shift of the order of 0.001 nm for each  degree of temperature change).  
More details on these filters are given in the following. 

  The recording device, which is placed in the front part of the telescope and above the objective lens is a  Xedar camera XS2048-A-12 equipped with a Thomson TH7899M 2048$\times$2048 pixels CCD sensor. The 14 $\mu$m$\times$ 14 $\mu$m pixels have a collecting capacity of 2$\times 10^5$ e$^-$ and a reading noise of 35 e$^-$. The camera has a dynamic range of 12 bit in reading of individual exposures and a reading speed of 4 frame/s, with the use of four amplifiers of different gain to read 
  the detector in  four 1024$\times$1024 pixels  quadrants\footnote{This can be seen very clearly in Fig. \ref{f4}.}.   
  The Xedar camera has been employed from September 1997 onward; the 
 Rome/PSPT prototype  used  a camera device based on a CCD Thomson (analog to digital ratio A/D at 12 bit/pixel) detector of 1024$\times$1024 elements. The lens in the secondary optics system that forms the focused beam on the detector, \textbf{is also} acting as entry window to the camera.
 
It is worth mentioning that, for many years, the CCD sensor of the Xedar camera  was cooled down  to 
-10$^{\circ}$C  during observations by a 
Peltier element, which was cooled by water.  However, the cooling and water flow were discontinued on September 2005, on the basis of the photometric characteristics of the acquired data. On the other hand, the Rome/PSPT camera has always been operated in light vacuum to prevent 
steam condensation on the CCD and to remove dust from it.

The pointing of Rome/PSPT  and motion of all its components have been controlled via simple control hardware placed within and near the telescope, and software written in C language and running on a PC 486; the data acquisition and telescope set-up for observations have been handled away from the telescope by codes written in C and IDL languages, which have been running on a Sun Ultra Sparc 1 workstation. The two computers have
 communicated through a RS-232 port, the Sun and PC assuming the roles of the \textit{master} and \textit{slave}, respectively.
The workstation has received the data from the CCD via a  optical-fibre cable, 
about 15 meters long.

\begin{figure*}
\begin{center}
\includegraphics[width=0.95\linewidth]{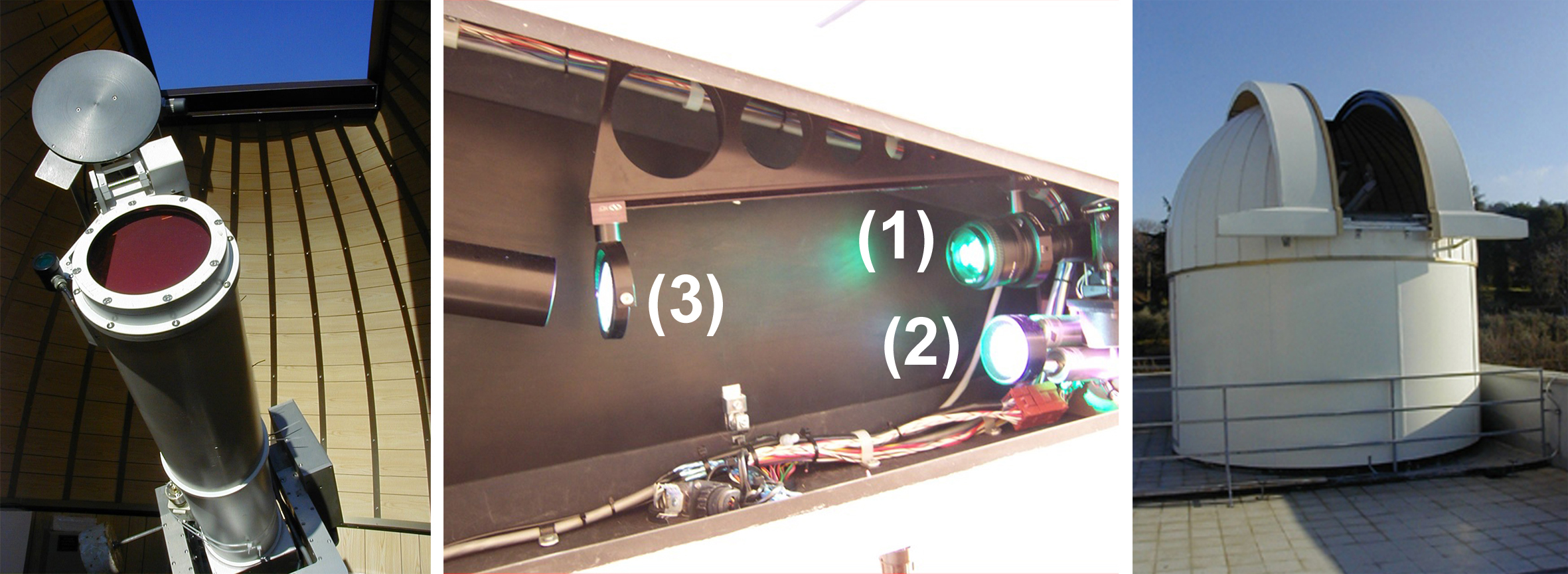}
\end{center}
\caption{Snapshots of the Rome/PSPT  in Monte Porzio Catone. From left to right: the objective lens  with its anti-infrared high-reflectance coating and dust cover; the guiding system with the quadrant-cell system (1), tip-tilt mirror (2), and beam-splitter (3); and the dome hosting the telescope. 
}\label{f2}
\end{figure*}

Figure \ref{f2} shows further images of  the Rome/PSPT, in particular of the objective lens (left panel) and guiding system (middle panel), and of the hosting dome (right panel). The latter, which has 3 m diameter, is placed on the roof of the institute, at about 10 m from the ground.

As mentioned above, in 1997, the prototype telescope installed at the Monte Mario site underwent a first update to a quasi-definitive version of the instrument. At that time, 
the body of the telescope, the electronics for the telescope monitoring and data acquisition, and the CCD detector were replaced with other components, whose main features are those described above.
Besides, the two interference filters  installed in the prototype instrument, centered at the blue continuum  at 409.4 nm (FWHM = 0.25 nm) and \ca line at 393.3 nm (FWHM = 0.25 nm),  were replaced with similar but better quality elements,  and a new filter was added for observations centered at the red continuum at 607.2 nm  (FWHM = 0.50 nm). Then, 
in September 2001, the Rome/PSPT underwent a second major update to meet the  characteristics of the MLSO/PSPT and NSOSP/PSPT. Besides, 
the telescope was moved to the Monte Porzio Catone site. At that time, 
the main instrumental changes consisted in the replacement of the 15-cm objective lens with a new one with better reflectivity, and the removal of a thermal filter included in the optical path up to that time.
Unfortunately, the upgrade of the control software to the version employed at the companion instruments could not be applied due to an incompatibility of local hardware. This incompatibility  is responsible for the different operational strategies applied with the Rome/PSPT and MLSO/PSPT. In April 2007, the Rome/PSPT was further  upgraded with two additional interference filters: one centered in the green continuum at 535.7 nm (FWHM = 0.50 nm) and one in the G-band at 430.6 nm (FWHM = 1.20 nm). In September 2008, the green continuum filter was replaced by  a new filter centred at the  \ca line at 393.3 nm  and characterized by a narrower bandpass (FWHM = 0.10 nm) with respect to the one of the filters employed in the telescope since its first-light. Table \ref{tab1} summarizes information of the interference filters used in the Rome/PSPT over the period 1996--2022. 

\begin{table*}
\caption{Interference filters installed in the Rome/PSPT. Listed are: filter name and label used to identify it, nominal transmission center and bandwidth (FWHM), period of operation.}  
\label{tab1}
\begin{tabular}{lcccc}          
\hline\hline                        
Filter & Label & Center & Bandwidth & Period\\
 &  & nm & nm & years	\\
\hline
\ca &K$_p$  & 393.3 & 0.25  &1996--1997\\
Blue cont & B$_p$ & 409.4 & 0.25 & 1996--1997\\
\ca & K  & 393.3 & 0.25  &1997--2022\\
\ca & KN & 393.3 & 0.10  & 2008--2022\\
Blue cont & B & 409.4 & 0.25 & 1997--2022\\
G-band & C & 430.6 & 1.20 & 2007--2022\\
Green & G & 535.7 & 0.50 & 2007--2008 \\
Red cont & R &607.2 & 0.50 & 1997--2022\\
\hline
\end{tabular}
\end{table*}

In addition to the changes described above, the Rome/PSPT suffered from failures of the electronics in the Xedar camera (from September to November 2004, and in August 2006) and in the \textit{slave} PC (August 2009) and  \textit{master} Sun (from May to December 2011, and in March 2022). All previous failures were recovered with fast and tailored extraordinary maintenance, except the most recent one that has been seen as an opportunity for a long overdue general update of the telescope to the technologies currently available.
Overall, the inactivity due to instrument maintenance has affected $\approx$ 3\% of the Rome/PSPT operational period from 1996 to 2022. 

Over the above interval, 
the Rome/PSPT  has delivered 
regular digital full-disk images of the Sun in the various spectral bands previously specified.  
The cadence of the acquired data has varied from one image per observing day to one image per 13 sec.  
Local sunshine and typical seeing conditions have  allowed observations of moderate spatial resolution on a routinely basis. In this respect, it is worth mentioning that the pixel size of 14 $\mu$m of the employed CCD in combination with the telescope optics correspond to $\approx$ 1 arcsec/pixel scale or an effective spatial resolution of $\approx$ 2 arcsec of the acquired data. However,  since early solar monitoring, Rome/PSPT observations have been  2$\times$binned during telescope operations, by accounting for typical seeing conditions at the observing site, which are often worse than 1 arcsec. 

The daily program of observations with the Rome/PSPT has included the acquisition of images obtained with each filter by summing up multiple exposures during each observation, usually 25 frames, and, from August 2000 onward, also of images from a  single exposure. All the observations have been taken by adapting the exposure time of the camera to reach suitable counts on the acquired images, without reaching their saturation. The exposure times have usually been in the range [30,60] ms  for each frame of all observations. 
Since early operations, the daily program of observations has also comprised acquisition of data for accurate instrumental calibration of the solar observations performed with the various filters, specifically dark-current and flat-field response data of the imaging device. Find details in the following. The acquisition of the latter data has required about 20 minutes for each filter employed, and about 2 hours for the five spectral bands regularly observed.

The Rome/PSPT observations have been generally carried out from about 8 UT to 13 UT, with best seeing conditions occurring in the interval 9--11 UT.  
On the other hand, the MLSO/PSPT observing day generally ran from about 16 UT to 02 UT, with best seeing conditions in the interval 16--19 UT.  
Therefore, on average, the observations obtained with the two telescopes have had a time difference between  about 4 and 18 hours.  
Normally a complete set of images from the various filters was acquired within a few minutes in order to have the data of the photosphere and chromosphere nearly simultaneously.  
Noteworthy,   
up to present, there have been only five different observers operating the Rome/PSPT. Moreover, the data taken from 2009 to 2022, i.e. during the last 14 years, have been obtained by the same  observer, 
who has also maintained the telescope and hardware components.

\begin{figure*}
\begin{center}
\includegraphics[width=0.95\linewidth]{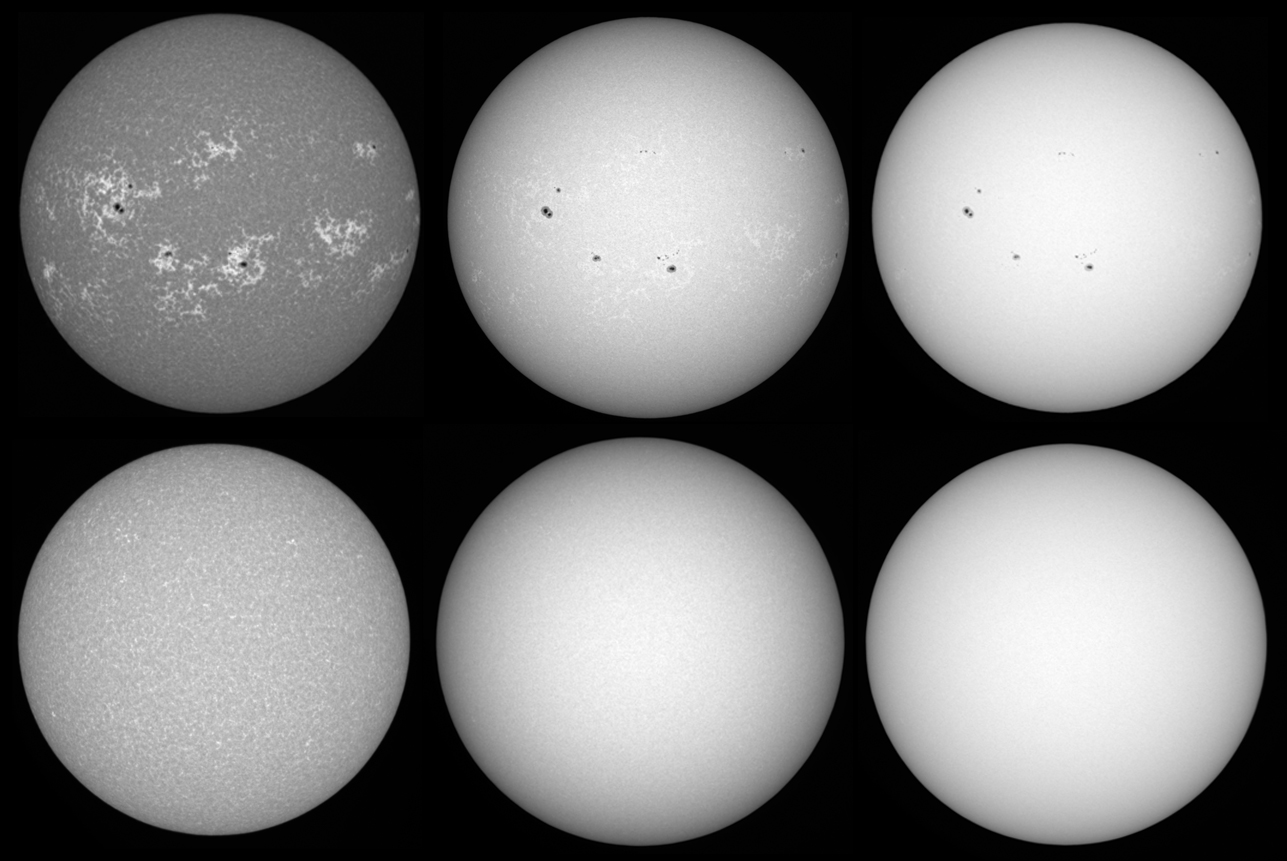}
\end{center}
\caption{Examples of the solar  full-disk images obtained with the Rome/PSPT on 13th December
2013 (top panels) and 30th June 2020 (bottom panels); from left to right: \ca at 393.3 nm (bandwidth 0.25 nm), G-band at 430.6 nm (bandwidth 1.2 nm), and  red continuum at 607.2 nm (bandwidth 0.5 nm) images. These images show examples of the routine Rome/PSPT observations after instrumental calibration. The solar disk is shown 
without compensation for ephemeris.   
}\label{f3}
\end{figure*}

Figure \ref{f3} shows examples of routine solar full-disk observations obtained with the Rome/PSPT at the \ca line at 393.3 nm (left panels), G-band at 430.6 nm (middle panels), and in the red continuum at 607.2 nm (right panels) on 13th December 2013 (top panels) and about six years later on 30th  June 2020  (bottom panels) at the maximum and minimum phases of solar cycle 24, respectively. A noteworthy aspect of Rome/PSPT images is that  the solar observation covers almost all the CCD sensor optimizing the pixel scale of the solar data.

\begin{figure*}
\begin{center}
\includegraphics[width=0.95\linewidth]{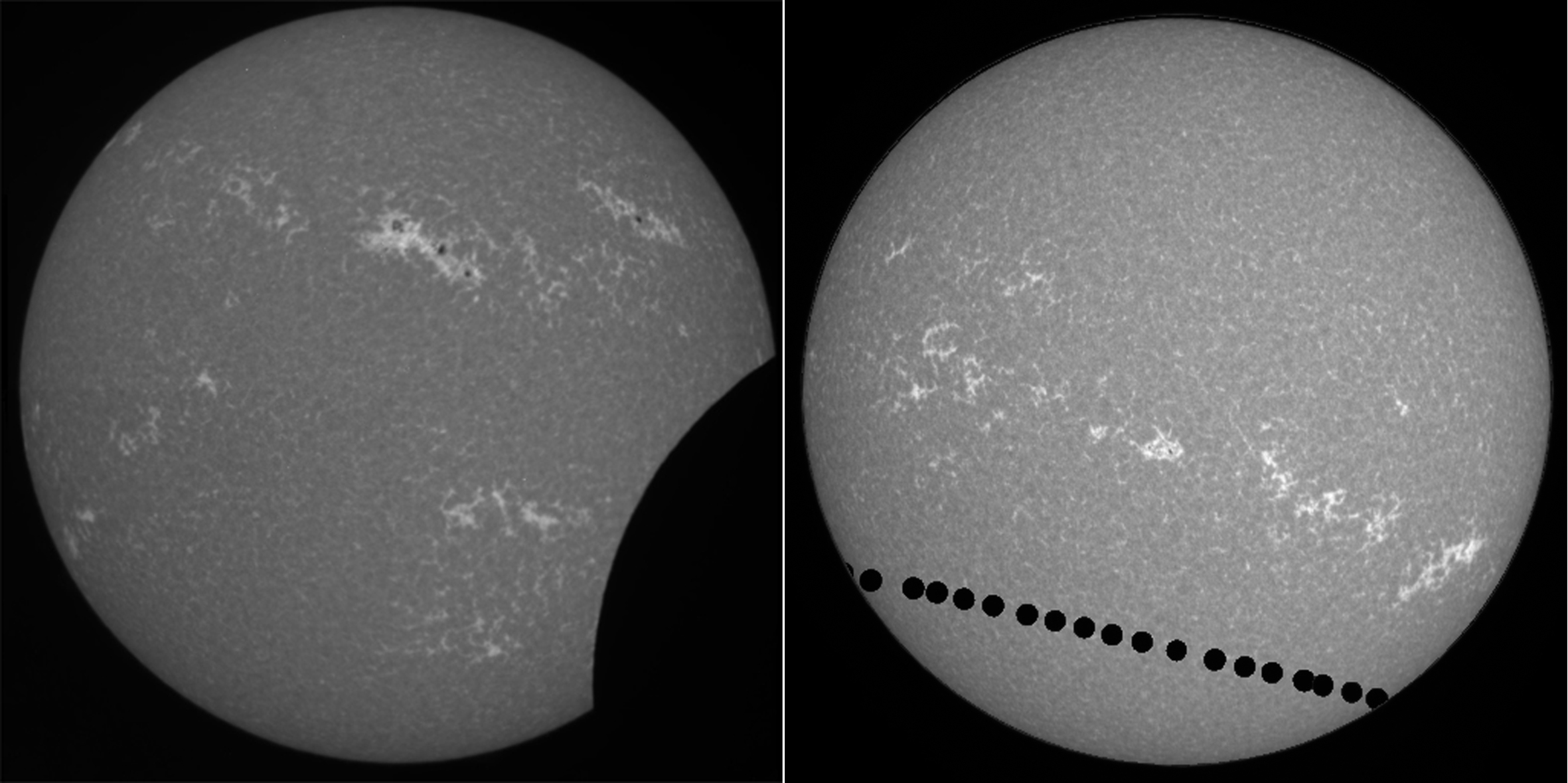} 
\end{center}
\caption{Examples of the solar  observations obtained with the Rome/PSPT at the \ca line during the partial solar eclipse on 11th August 1999 (left panel) and transit of Venus on the solar disk on 8th June 2004 (right panel). The latter image is a composite, which was produced post-facto by stitching together the best observations of the Sun during Venus' transit. 
}\label{fecl}
\end{figure*}

Figure \ref{fecl} shows examples of the occasional observations acquired during astronomical events, e.g. the partial solar eclipse in August 1999 (left panel) and the transit of Venus on the solar disk in June 2004 (right panel). In addition to serve public outreach activities, these observations were exploited for instrumental calibrations, specifically to get estimates of the stray-light and of the  \textit{Point Spread Function} (PSF) of the Rome/PSPT.

\section{Photometric calibration}

\begin{figure*}
\begin{center}
\includegraphics[width=0.95\linewidth]{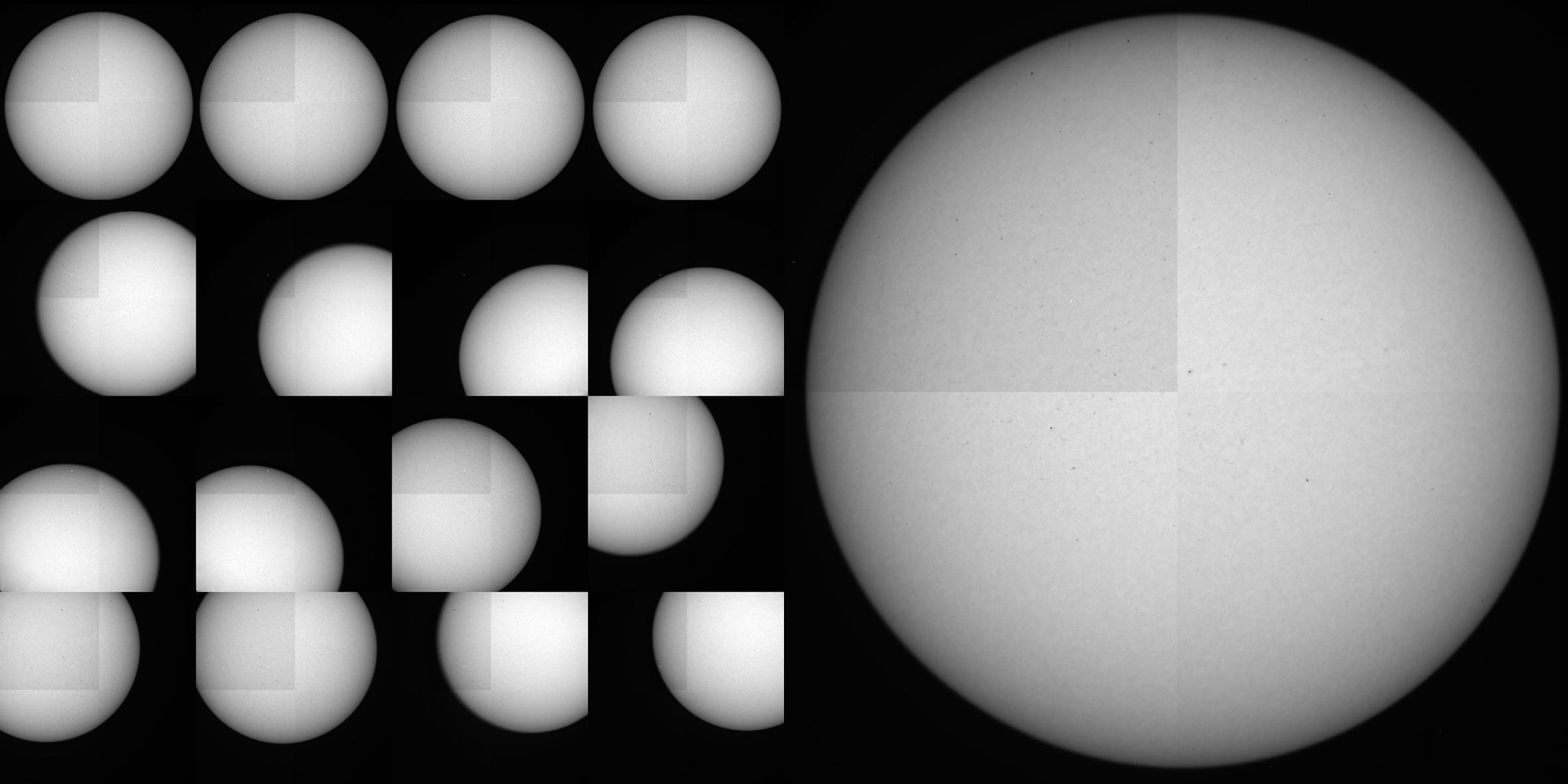} 
\end{center}
\caption{Examples of the set of displaced images (left panel), each sum of 10 exposures, obtained for the flat-field calibration of the Rome/PSPT blue continuum observations taken on 27th April 2018 and reference image (right panel), sum of 25 exposures, centred on the CCD device. The images in the two panels are shown with different size for the sake of ease.  The slightly different gain of the four amplifiers employed to read the CCD detector is easily noticed in all the   images.  
}\label{f4}
\end{figure*}

\begin{figure*}
\begin{center}
\includegraphics[width=0.95\linewidth]{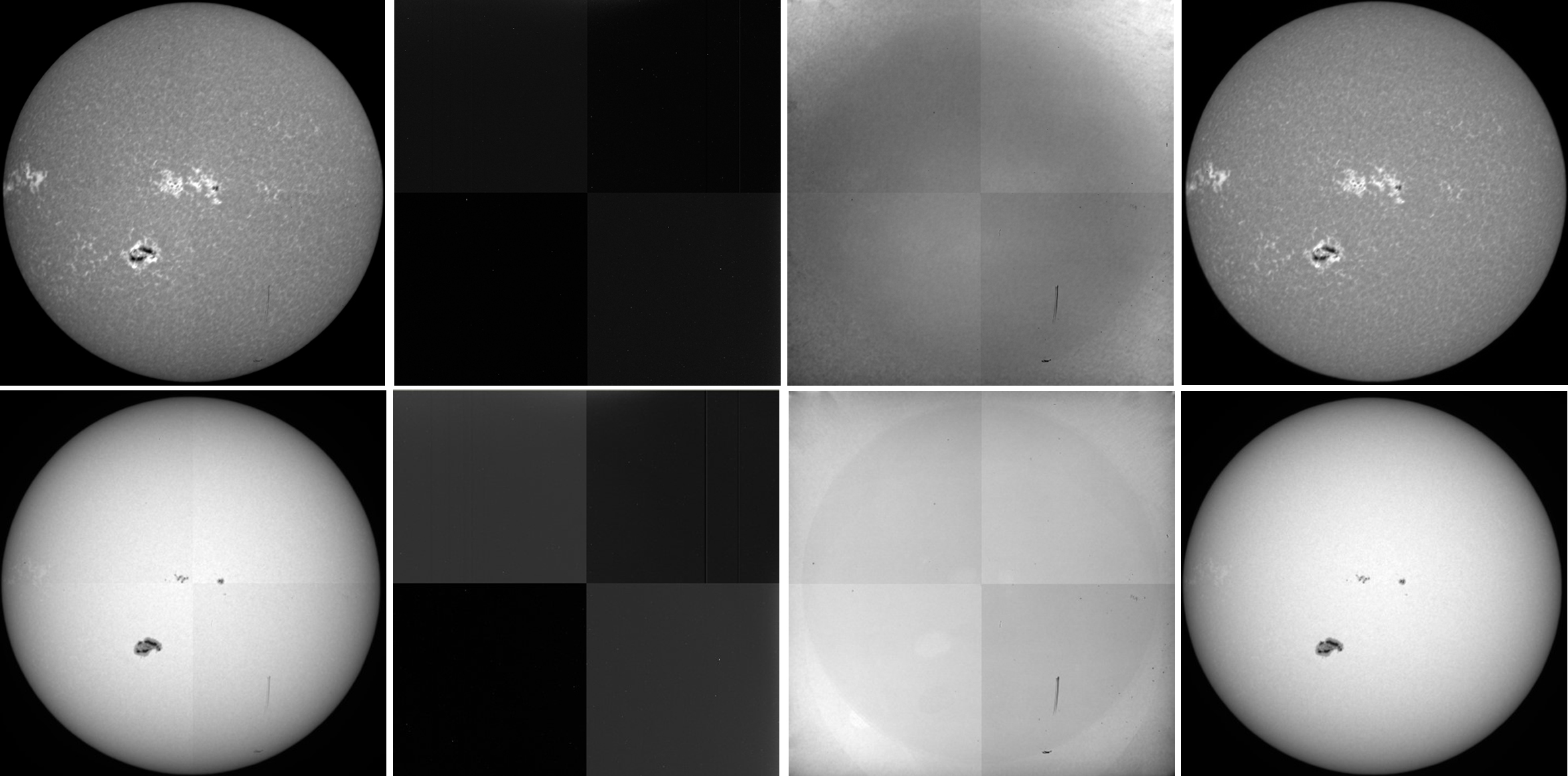} 
\end{center}
\caption{Examples of the raw and calibrated \ca (top panels) and blue continuum (bottom panels) observations obtained on 14th January 2005, and relative calibration data.   
From left to right in each row: raw image, dark frame, flat-field response, and calibrated image. The quadrant pattern in raw observations and calibration data generated by the four amplifiers of the recording camera is accurately compensated for in the calibrated observations, as are the small-scale dark structures in the raw observations and flat-field images produced by permanent inhomogeneities of the substrate of the detector and occasional dust and hairs on it.  
}\label{f5}
\end{figure*}

In addition to careful instrumental design,  
 high-precision photometry requires accurate  calibration of the data acquired. The latter mostly implies the bias and dark-current frame compensation, and precise  characterization of the response of the recording device to a flat-field illumination. 
 
  Since early Rome/PSPT observations, the bias and dark-current frames have been taken with the exposure times of each solar observation at shutter closed and unexposed CCD, while the flat-field response frames have been estimated with the \textit{displaced-images} method proposed by \citet[][]{kuhn1991}. This method  
uses a set of Sun images displaced on the CCD to recover the gain of individual pixels of the sensor via iterative computations of the intensities recorded at each pixel in the displaced images. 
The telescope is unchanged during the acquisition of the multiple displaced images with respect to the regular solar monitoring. Under this condition the method allows the estimation of the spatial non-uniformity of both the electronic gain of individual pixels of  the CCD camera and of the radiation beam upon it due to the optics of the telescope. This method is specifically suited for  the flat-field calibration of images of extended sources like  solar full-disk observations.

In principle, the \textit{displaced-images} method enables the detection of relative changes of pixels response on the CCD  with an accuracy higher than 10$^{-4}$. However, its  application is founded on several   
assumptions that are only partly realized with the real observations. These 
assumptions are: 
1) the CCD detector has a linear response;
2) the gain inhomogeneities derived from the method account for both the pixel-to-pixel changes of the electronic gain of the CCD sensor and other multiplicative contributions to the CCD response due to the optics of  the telescope;
3) the Sun does not change its appearance during the acquisition of the displaced images; 
4) the solar disk center is known without error in all the displaced images. 
While assumptions one and two are well accomplished during observations, hypotheses three and four are usually not fully realized, due to the evolution of the  atmospheric seeing and of solar regions, and uncertainties  
in detecting the solar limb  
under some conditions.    
Tests performed on synthetic images have shown that the nominal accuracy of the method can be reached only if the disk center of the displaced images is determined with an accuracy better than 1 pixel in all the employed data, and by using displaced images that cover the CCD sensor homogeneously with both small and large image shifts \citep[e.g.][]{ermolli1998,ermolli2003}. Lack of these conditions result in flat-field response frames showing  spurious \textit{ghost images} of the solar limb and of the features observed on the solar disk. In these conditions, 
the accuracy of the flat-field response determined with the method 
decreases up to 10$^{-2}$, as in the case of solar images
acquired under poor observing conditions or with inadequate data.

After 
several optimization tests performed on real and synthetic images, the \textit{displaced-images} method has been applied to Rome/PSPT observations with a set of 17 solar images.  
For each spectral observation, these images have consisted of 16 observations   obtained by shifting the solar disk center from 5 to 150 pixels with respect to the image center, and a reference solar image centered upon the CCD detector. Each of the 16 displaced images has been obtained summing up 10 consecutive exposures,  while the reference image by adding 25 exposures. The image displacements have been applied 
in the vertical and horizontal direction to cover both small- and large-displacements of the solar image, as suggested by the tests performed on synthetic data.  A  
dark-current frame has complemented the above data set for its proper processing. The multiple displaced data have been recorded for each filter once a day.

Figure \ref{f4} shows an  example of the 16 displaced images and reference frame acquired for the calibration of the Rome/PSPT observations taken in the blue continuum on 27th April 2018. Each image in the set shows the quadrant pattern typical of any Rome/PSPT raw observation, due to the four amplifiers with   different gain used for the reading of the image data. 
Notwithstanding the limitations described above, the tests performed with Rome/PSPT data have shown that under stable observing conditions the accuracy of  flat-field calibration obtained with the \textit{displaced-images} method applied as specified above  largely exceeds the one achieved with other methods, by using e.g. an uniform light diffuser in the optical path or multiple images of the sky. Also at the MLSO/PSPT  the \textit{displaced-images} method was applied by using 16 off-set images of the Sun \citep[e.g.][]{rast2008}.

In the Rome/PSPT, the acquisition of the  displaced images has been performed with the computer-control software employed for regular solar observations. The flat-field response frame has been derived from the set of displaced images acquired with each filter by using a procedure written in IDL language and  
partly based on codes written by \citet{kuhn1991}. 

Figure \ref{f5} shows an example of the Rome/PSPT raw (left panels) and calibrated (right panels) images from observations in the \ca (top panels) and blue continuum (bottom panels), and their respective calibration data (middle panels), the dark-current and flat-field response frames.
The flat-field images show large-scale intensity anisotropies due to the diverse amplifiers of the camera and telescope's optics ranging [2,12]\% in the chromospheric \ca data and  [1,8]\% in the  photospheric continuum and G-band data. The pixel-to-pixel differences in the flat-field images range [6,12]\% in the  \ca data and [5,9]\% in the continuum and G-band data. 

All the Rome/PSPT observations have been processed for dark-current and flat-field calibration. 
Observations acquired with unfavorable weather conditions that prevented the acquisition of calibration series have been processed by using a flat-field response computed from same band close in time observations. Observational data show that flat-field images slightly differ from one day to the other, likely due to small differences in the  observing conditions and accuracy of the flat-field image computation. However,   flat-field images derived  from observations taken with all filters over a few days and a period of eleven years differ on average by less than  5$\times$10$^{-4}$ and 10$^{-3}$, respectively.

\section{Image quality}

\begin{figure*}
\begin{center}
\includegraphics[width=0.95\linewidth]{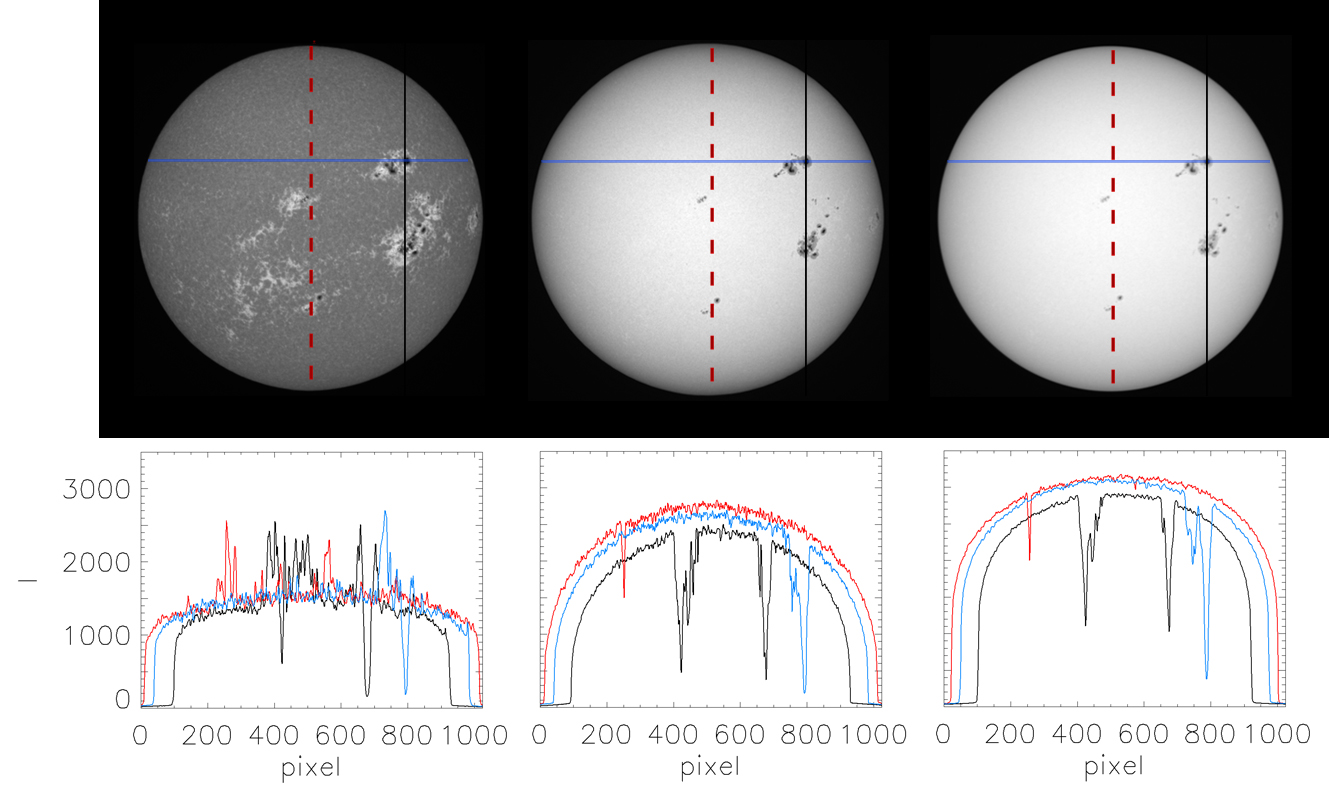}
\end{center}
\caption{\textit{Top: }Examples of the solar full-disk  observations acquired with the Rome/PSPT on 31st October 2003 at the \ca line (left), blue continuum (middle), and red continuum (right). 
\textit{Bottom: } Intensity profiles over linear segments (marked in the same color in the full-disk images) derived from the solar observation directly above them. 
}\label{f6}
\end{figure*}

\begin{figure*}[ht!]
\begin{center}
\includegraphics[width=0.95\linewidth]{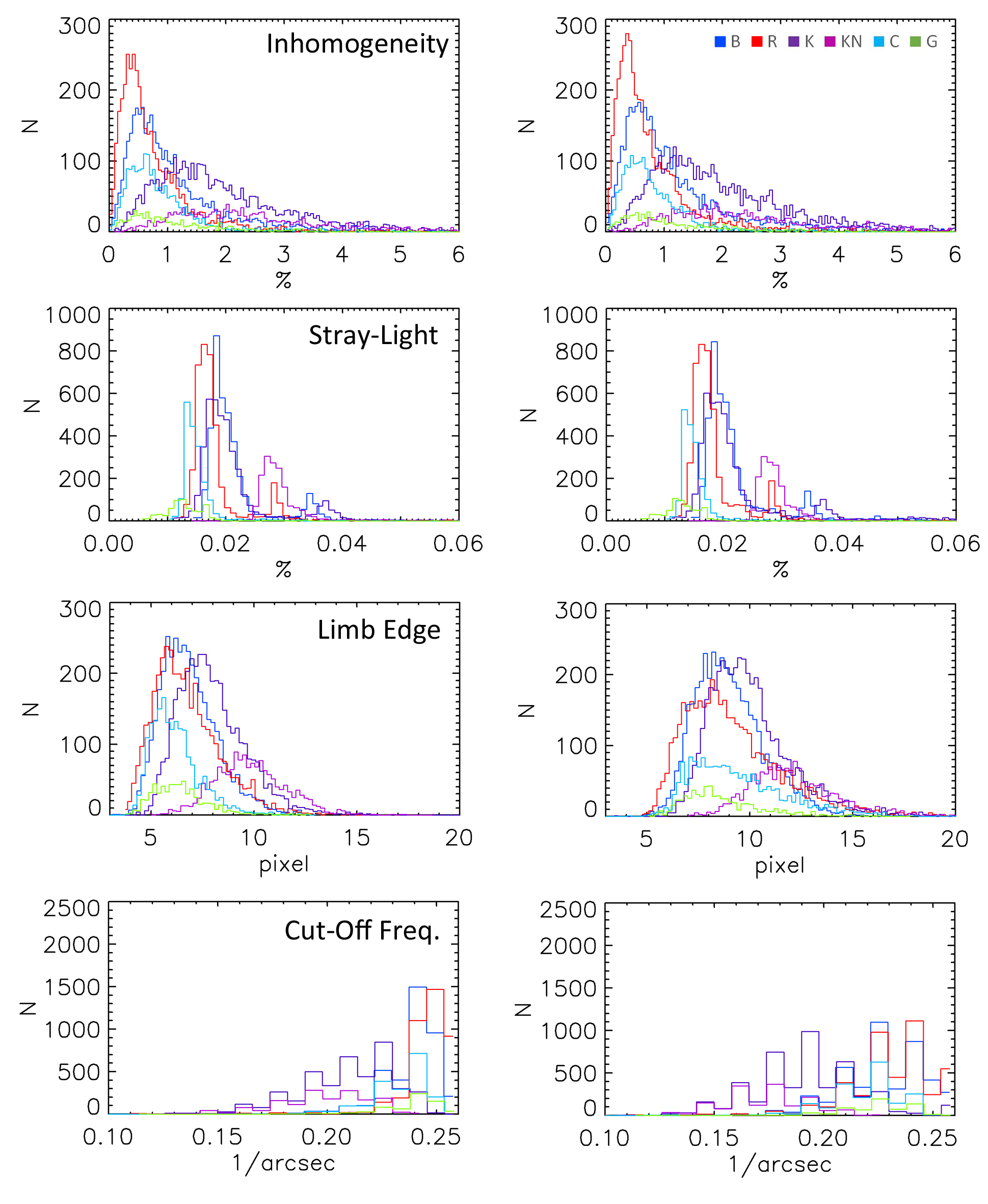} 
\end{center}
\caption{Distribution of the values of the parameters measured in each Rome/PSPT image to characterize its quality, for  single exposure (left panels) and sum of 25 exposures (right panels) observations. From top to bottom each row displays: level of large-scale intensity inhomogeneity at $r=0.5R_{\mathrm{Sun}}$ [\%], stray-light level at $r$=1.05--1.07 $R_{\mathrm{Sun}}$ [\%], FWHM of the solar limb edge [arcsec], and cut-off frequency  detected [1/arcsec], for observations taken with the diverse filters as specified in the legend. 
}\label{fqua2}
\end{figure*}

\begin{figure*}
\begin{center}
\includegraphics[scale=0.95,trim=0 0 0 250 ,clip]{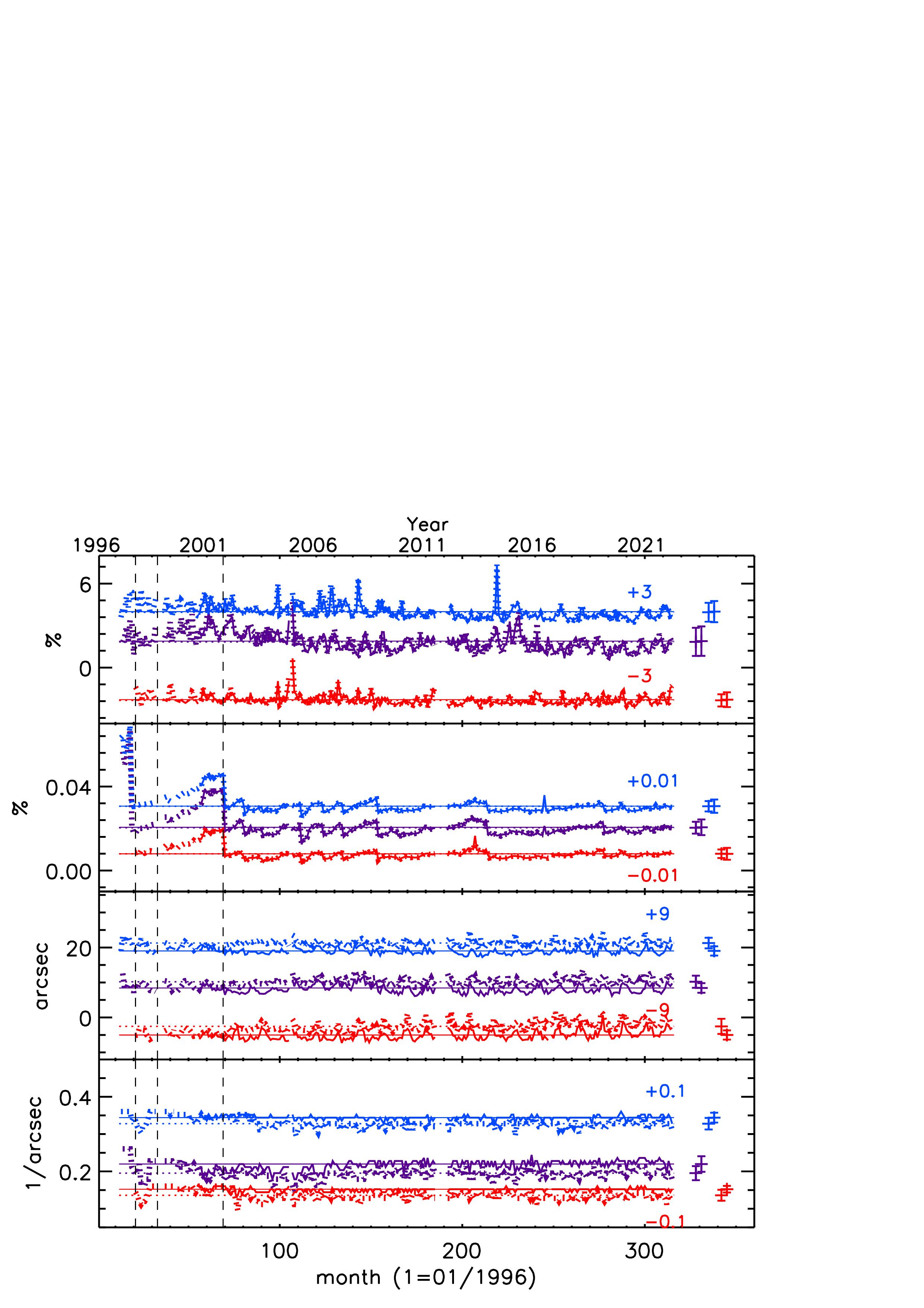} 
\end{center}
\caption{Evolution of monthly median value of the parameters measured to characterize the quality of the sum of 25 exposures (dotted lines) and single  exposure (solid lines) observations. From top to bottom: level of large-scale intensity inhomogeneity at $r=0.5R_{\mathrm{Sun}}$ [\%], stray-light level at $r$=1.05--1.07 $R_{\mathrm{Sun}}$ [\%], FWHM of the solar disk edge [arcsec],  and cut-off  frequency detected [1/arcsec] in the data taken with the \ca wider band (K; violet line), blue (B; blue line), and red (R; red line) filters. Vertical dashed lines mark times of instrumental discontinuities. Horizontal dashed and solid lines show the mean of the values measured on sum of 25 exposures and single exposure observations, respectively. For the sake of clarity, B and R data are shifted by the values specified in each panel. 
 See Section 4 for more details. 
}\label{fqua}
\end{figure*}

\begin{figure*}
\begin{center}
\includegraphics[scale=0.95,trim=0 0 0 250 ,clip]{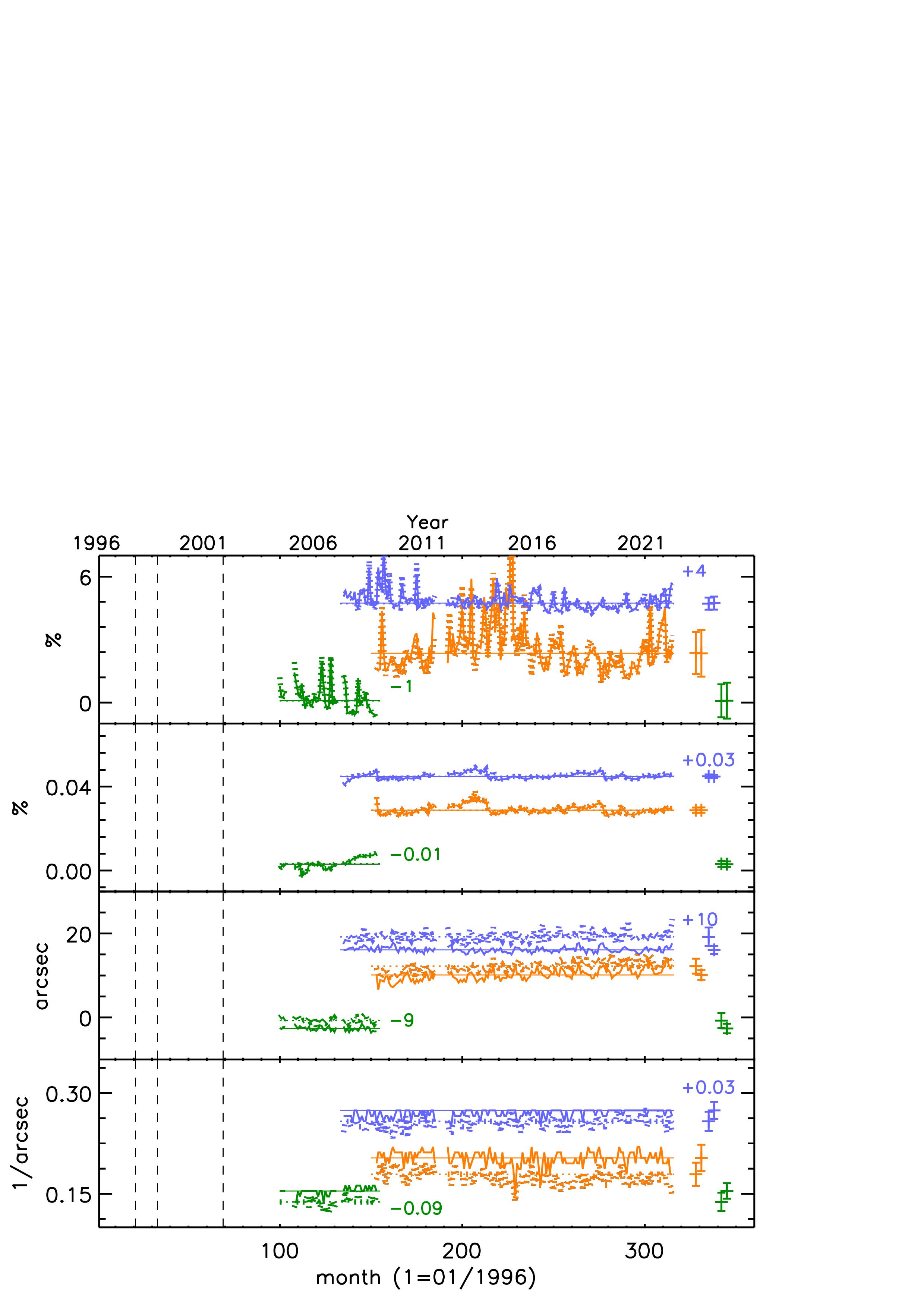} 
\end{center}
\caption{Same as Figure \ref{fqua} but for observations  taken with the \ca narrower band (KN; orange line), green continuum (G; green line), and G-band (C; violet line) filters. Vertical dashed lines mark times of instrumental discontinuities. Horizontal dashed and solid lines show the mean of the values measured on sum of 25 exposures and single exposure observations, respectively. For the sake of clarity, G and C data are shifted by the values specified in each panel. 
}\label{fqua3}
\end{figure*}

Figure \ref{f6} shows examples of calibrated images (top panels) obtained from the Rome/PSPT observations  acquired on 31 October 2003 at the \ca line (left panel), blue continuum (middle panel), and red continuum (right panel).  Two very large sunspot groups were present on the solar disk on that day, as well as several bright facular regions seen as plages in the \ca image and just close to the limb in the blue continuum data.  
The profiles plotted below each full-disk observation (bottom panels) trace the intensity counts along the horizontal (blue) and vertical (red and black) lines  overplotted to the solar image just above.  The regularity of these profiles suggest the lack of any severe large-scale image anisotropies in addition to a linear gradient in the intensity counts,     
which is spotted on, e.g. the vertical profiles over the \ca and red continuum observations. This gradient is due to differential atmospheric extinction over the extended solar disk observed with the telescope. The amplitude  of this intensity gradient depends on the  transparency of the atmosphere during observations, elevation of the observing site, and Sun's zenith angle. In Rome/PSPT data the measured maximum intensity gradient is lower than 1\% and in the range [0.3,0.9]\%.   
Faculae, sunspots and other smaller-scale brightness features on the solar disk, such as the network pattern, are readily apparent in the \ca images with a change of the intensity values larger than $\approx$ 50\%, 50\%, and 10\% for faculae,  sunspots, and network regions, respectively, the former exceeding the  
variation of intensity values over most of the solar disk. The blue and red continuum images clearly show umbral and penumbral regions of sunspots and smaller-scale bright and dark features at a somewhat different spatial pattern and with a change of the intensity values of up  $\approx$ -80\% and +5\% for sunspot and other disk features, respectively.

In addition to the regular check of the  intensity values in the images during and after the data acquisition,  
since early observations the quality of the Rome/PSPT images has been studied post-facto by using their properties  \citep{fazzari2003}.
In particular, various parameters have been derived from each calibrated image to characterize its quality in both global  and local scales, by considering the level of large-scale intensity inhomogeneities, of stray-light, and of sharpness in the image. 

As mentioned above, large-scale intensity inhomogeneities in the image are intrinsically due to differences in the atmospheric extinction over the observed solar disk, which produce an intensity linear gradient in the vertical direction in the image. In addition, solar full-disk  observations of the photosphere show the typical limb-darkening and center-to-limb intensity variation due to the decrease of solar plasma temperature with height in the observed atmosphere. The limb-darkening also characterizes the chromospheric observations at the \ca obtained with relatively wide filters as the ones employed in the Rome/PSPT, which include a significant source contribution from the photosphere.  However, intensity inhomogeneities in the images can be additionally due to e.g. telescope's optical set-up, inaccurate instrumental calibration, and clouds during observations. The level of such intensity anisotropies has been estimated in all the acquired Rome/PSPT data by measuring the maximum difference in the average intensity computed along rings centered on the solar disk and  at various distances from the disk center. In the raw images taken under clear sky conditions, the average intensity computed along a ring at e.g. $r=0.5R_{\mathrm{Sun}}$ distance from disk center, where $R_{\mathrm{Sun}}$ is the solar disk radius, and over the four quadrants of the recording device can differ up to 12\% due to the diverse characteristics of the four amplifiers employed in the recording device.  
Standard instrumental calibration applied to the data and usual observing conditions allow the above difference to decrease to maximum values in the range  $\approx$ [0.5,0.9]\% for the observations performed with the continuum and G-band filters, and  in the range $\approx$ [1.7,2.2]\% for those with the \ca filters. On the other hand, inaccurate calibration and unfavorable weather conditions can increase the maximum intensity anisotropy in the image up to 3--5\%. It is worth recalling that the above maximum values include contributions from small-scale disk features outside active regions and from the intensity linear gradient in the images.

The level of stray-light in the images derives from characteristics of the telescope and site of its operation. In Rome/PSPT observations, this level has been estimated by measuring the intensity in the solar aureola at 1.05--1.07$R_{\mathrm{Sun}}$  distance from disk center, normalized to the disk center intensity. Results of such measurements show median values lower than 0.02\% with standard deviation values in the range [0.006,0.012]\%, thus suggesting that the Rome/PSPT data are only minutely affected by stray-light, except for the ones taken with the narrower \ca filter. Indeed, those images  seem to suffer from a slightly higher stray-light degradation than other data, being the median of the measured intensity 0.029\% with standard deviation value 0.006, likely due to a  
reflection of the optical beam on filter's surfaces or other optical elements. Overall, the stray-light contribution estimated in the Rome/PSPT data mostly account for the term due to the atmosphere, which is typical of an observing site at low elevation. 

The image sharpness is affected by atmospheric seeing during observations and characteristics of the guiding system of the telescope, as well as  observational procedures.   
The sharpness of Rome/PSPT images has been evaluated by measuring the width of the solar disk edge in each observation and the smallest spatial scale detected on it. The edge width has been derived as the full-width-half-maximum (FWHM) of the Gaussian function fitting the intensity gradient on the disk edge. The smaller scale detected in the image has been derived from the cut-off frequency in  the power spectrum of the Fourier  bi-dimensional transform of a central sub-array of each observation. Results of both these measurements suggest that the median residual image blurring affecting the observations ranges  $\approx$
[6,12] arcsec, while the median of the cut-off frequency corresponds to smallest scale detected is in the range $\approx$ [4.0,5.2] arcsec.

Figure \ref{fqua2} shows the distribution of values derived for each of the above parameters from their measurements  on the observations obtained with single exposure (left panels) and sum of 25 exposures (right panels). Comparison of results achieved from the two classes of observations shows that the large-scale intensity anisotropy and the stray-light in the images are unaffected by the sum of exposures during observations.  
On the other hand, single exposure data are characterized by a slightly higher sharpness than the observations derived from the sum of 25 exposures; the average and standard deviation of the FWHM of the disk edge (smallest scale detected)  are 
6.4$\pm$1.4 (4.1$\pm$0.1) arcsec and 
8.8$\pm$2.0 (4.8$\pm$0.1) arcsec for the single exposure continuum and  \ca data, and 
8.7$\pm$2.1 (4.3$\pm$0.3) arcsec and 
10.9$\pm$2.5 (5.1$\pm$0.3) for the 25 exposures corresponding  data, respectively. 

Figures \ref{fqua} and \ref{fqua3} show the evolution of the monthly mean of the above parameters measured on the whole series of Rome/PSPT observations, in order  to quantify their homogeneity over time. 
In particular, Figure \ref{fqua} displays the results derived from the longest series of observations available, which are the ones obtained with the \ca, blue continuum, and red continuum filters, while Figure \ref{fqua3} shows the findings from the other shorter series of observations, at the green continuum, G-band, and \ca with narrower band. In both Figures, dotted and solid lines describe results from observations obtained from sum of 25 exposures and from single exposure, respectively.   
For the sake of clarity,   
to allow distinguishing results from the diverse series, the values derived from some data sets   
have been shifted by the quantities specified close to the values plotted for those data.
Vertical dashed lines mark times of instrumental discontinuities. Horizontal dotted and 
solid lines show the median value over the whole period for observations sum of 25 exposures and 
single exposure, respectively. Bars indicate median of the quantity and of its standard deviation over the whole data series.

All the parameters in Figures \ref{fqua} and \ref{fqua3} display changes on time scales from a few months to longer-term intervals, which will be explained in the following.

The large scale inhomogeneities measured in the images exhibit a slightly decreasing trend for all the data acquired. However, the results from the images taken with the  
\ca filters also display  an increase of the anisotropy measured in the data acquired  during activity maxima, with peaks at years 2000, 2002, and 2014. Images taken with the narrower  \ca filter manifest a greater increase of the measured anisotropy than that estimated for the data obtained with the wider \ca filter. It is worth 
mentioning that the roughly decreasing trend of  large scale inhomogeneities of sum of 25 exposures Rome/PSPT \ca  observations was also reported by \cite{chatzistergos2019sp}, from a different methodology. The nature of the decreasing trend is unclear and currently under investigation.

The level of stray-light  measured in the images is rather constant  
for all the data acquired since 2001. As clearly seen in  Figure \ref{fqua}, the update and relocation of the telescope occurred in 2001 led to a significant decrease of the stray-light affecting the data. For all the data acquired, there are clear and steady increases in stray-light estimates, which are discontinued at times of telescope's optics cleaning.
Results in both Figures  clearly show that the difference between values of the intensity anisotropy and stray-light measured from sum of 25 exposures and single exposure observations are minor, as the deviation of values over the whole interval of available data.  On the other hand, the values of the disk edge FWHM derived from all the data show a clear annual change. This is not surprising due to the varying observing conditions over the course of the year, because of  the annual change of the Sun's image on the CCD sensor and of the seeing conditions.
Besides, for all the data and for both the estimation of the disk edge FWHM and cut-off frequency in the image, the values obtained 
from observations that are sum of   25 exposures are slightly worse than the ones measured on single exposure observations. This is not surprising considering that the latter data are taken over a shorter time interval than the sum of 25 exposures observations. Moreover, results of the disk edge FWHM  
from the former set of observations  
exhibit a slight increasing trend in all the data   analysed. This increase is however minute with respect to the annual deviation of the  measured values,  
but for the ones obtained with the narrower \ca filter. The above slight increase is also observed in the values derived from single exposure images, but it is even less  pronounced with respect to the one derived from the data sum of exposures, yet with the exception of narrow \ca images. It is worth noting that the observed trends could be due to instrument aging, as well as to changes in the observing conditions. In the lack of ambient data at the telescope, relevant e.g. to temperature and humidity of air during observations, planned measurements of the filters' transmission may provide insights on the nature of the observed trends. 

The cut-off frequency  detected in the images, assumed as a measure of the smallest scale observed therein is rather constant for all the data. Single exposure observations are often limited by the pixel scale of the images, while observations obtained from sum of 25 exposures display a slightly lower cut-off frequency. 
Slightly smaller scales are detected in the \ca observations obtained since 2004.

It is worth mentioning that the above parameters studied in Rome/PSPT observations since early telescopes' operations, or other quantities derived from these parameters,  have also been used to characterize the quality of MLSO/PSPT images \citep[e.g.][]{rast2008} and of other historical and modern full-disk observations  \citep[e.g.][]{ermolli2009,chatzistergos2019sp,potzi2021}. Noteworthy, the estimation of the above parameters conducted on MLSO/PSPT images gave comparable results to those achieved from Rome/PSPT data, but for the stray-light level and sharpness of continuum observations, in favour of the MLSO/PSPT data. This is unsurprising due to the large difference of elevation of the observing sites of the two telescopes. On the other hand, all the parameters for the \ca data are in favour of the Rome/PSPT series, most likely due to transmission properties of the filters and other optical elements employed in the two telescopes.  Unfortunately, the stable solar monitoring by the MLSO/PSPT only extends the 2005--2015 period, thus over 
roughly one third of the whole operational interval of the Rome/PSPT. Nevertheless, the twin facilities can help distinguish between instrumental contributions and solar signatures 
in the evolution of any quantities measured in the observations taken over the common observing period.  
On the other hand, there are a few other archives of solar  full-disk  observations performed  over the last decade that can help singling out instrumental effects on Rome/PSPT observations, as done e.g. for the construction of the composite plage area evolution over 1892--2019 by \citet[][]{chatzistergos_analysis_2020}.

\section{Data archive}

\begin{figure*}
\begin{center}
\includegraphics[width=0.95\linewidth]{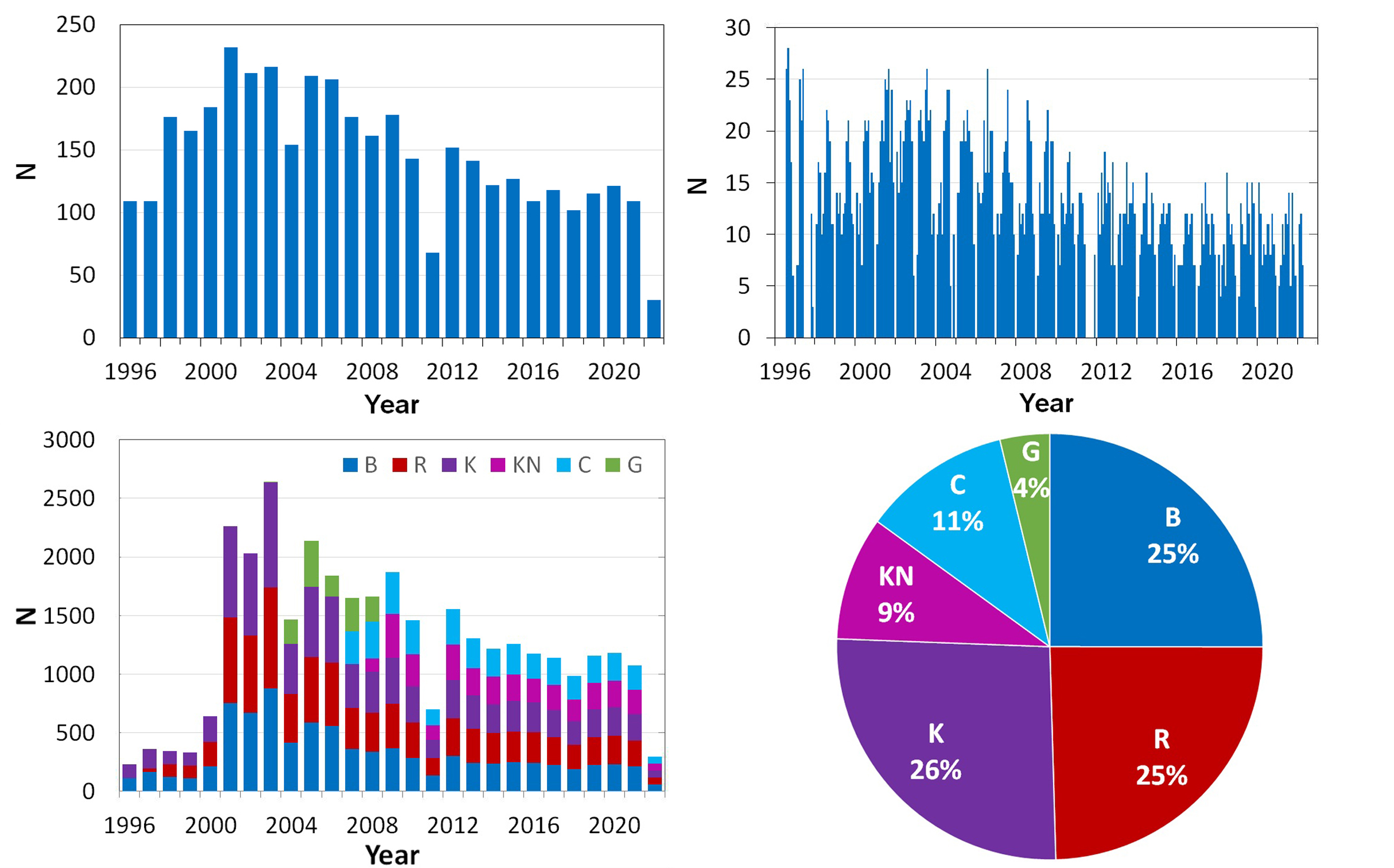} 
\end{center}
\caption{\textit{Top: }Number (N) of observing days per year (left panel) and month (right panel) from 1996 to 2022. \textit{Bottom: }Number (N) of Rome/PSPT observations at the diverse spectral bands obtained per year from 1996 to 2022  (left bottom panel) and statistics of the data available at the diverse observed bands (right bottom panel)
}\label{f10}
\end{figure*}

The Rome/PSPT project aimed at acquiring accurate photometric observations of the solar  full-disk photosphere and chromosphere at various spectral bands over long time spans. 
This has been done regularly since July 1996, but with three slightly different telescope set-ups. These have resulted in short  discontinuities in the telescope's operation in  September 1997 and 2001, the latter to update the facility to final  characteristics.

 In Figure \ref{f10} (top panels)  we show the total number of 
observing days per year (left panel) and per month (right panel) from 1996 to 2022. It can be noticed that the number of observing days has increased as time went on until year 2001 when the Rome/PSPT monitoring activity started to suffer from reduction of telescope operation on 5/7 days per week; the number of observing days then further started a progressive decrease after 2005, when the monitoring was also subject to the availability of  the sole observer. The number of observations per month has followed typical seasonal trends of local weather and annual leaves; months with just a few or without observations in summer 1997, fall 2001, and in year 
2011 are due to the first instrument upgrade, the further upgrade and relocation to Monte Porzio Catone site, and camera failure described in Section 3.

The observations recorded over the 27-year long period of regular Rome/PSPT operations have created a significant data archive consisting of more than 30000 images of the full-disk photosphere and chromosphere and 1,5 Tb of data, 50\% of which consisting of calibration series. Moreover, the Rome/PSPT is one of the few solar full-disk facilities currently operated, and which have made daily observations  overlapping the historical time-series of data obtained  with spectroheliographs at e.g. the Meudon, Coimbra, Kislovodsk, and Kharkiv observatories \citep[e.g.][]{chatzistergos2022}.

Most of the daily routine observations have produced one calibrated image per each observed spectral band  from both single exposure and sum of 25 exposures frames. Since the former observations started in 2001 only, the Rome/PSPT data archive comprises single exposure and sum of 25 frames full-disk solar images for 48\% and 52\% of the available data, respectively. In addition to these routine observations, the telescope has been set for dedicated observing campaigns, for high cadence observations at a given band, during astronomical events as e.g. solar eclipses (in 1999, 2006, 2015)  and planet transits (Mercury and Venus in 2003 and 2004, respectively), and for parallel observations with other instruments. These data however represent only 5\% of the available observations.

In Figure \ref{f10} (bottom panels) we show the number of Rome/PSPT observations at the diverse spectral bands (left panel) obtained per year from 1996 to 2022 and statistics of the data available at the diverse observed bands (right panel). The observations acquired per year with all the filters and the final version of the instrument have been on average 
146, with the number ranging from 68 in 2011 and 232 in 2001. The observations taken at the \ca with the broader filter and at the blue and red continua represent the 26\%, 25\%, and 25\% of the available data, respectively, while observations at the \ca taken with the narrower filter available, at the G-band and at the green continuum represent the 11\%, 9\%, and 4\% of the Rome/PSPT data.

The observations have been calibrated at the end of each observing day and calibrated data have been made available to the community with just a few hours delay due to the processing of the calibration data. The Rome/PSPT archive consists of all the raw solar images and calibration data stored as Flexible Image Transport System \citep[FITS;][]{wells_fits_1981} files from 1996 to 2022 and corresponding calibrated observations also stored as FITS files and as JPG quick look files. The latter can be accessed at the following link \url{https://www.oa-roma.inaf.it/pspt-daily-images-archive/}  
that has been updated regularly. 

The Rome/PSPT data archive represents a unique resource for studies of the solar variability and activity.  Indeed, it has monitored the Sun  change over two  complete solar cycles - from very quiet to very active, and back again, this twice, and even more. Also noteworthy is that the Rome/PSPT data archive  includes such information in both the photosphere and chromosphere.

\section{RESULTS}

\begin{figure*}
\begin{center}
\includegraphics[width=0.95\linewidth]{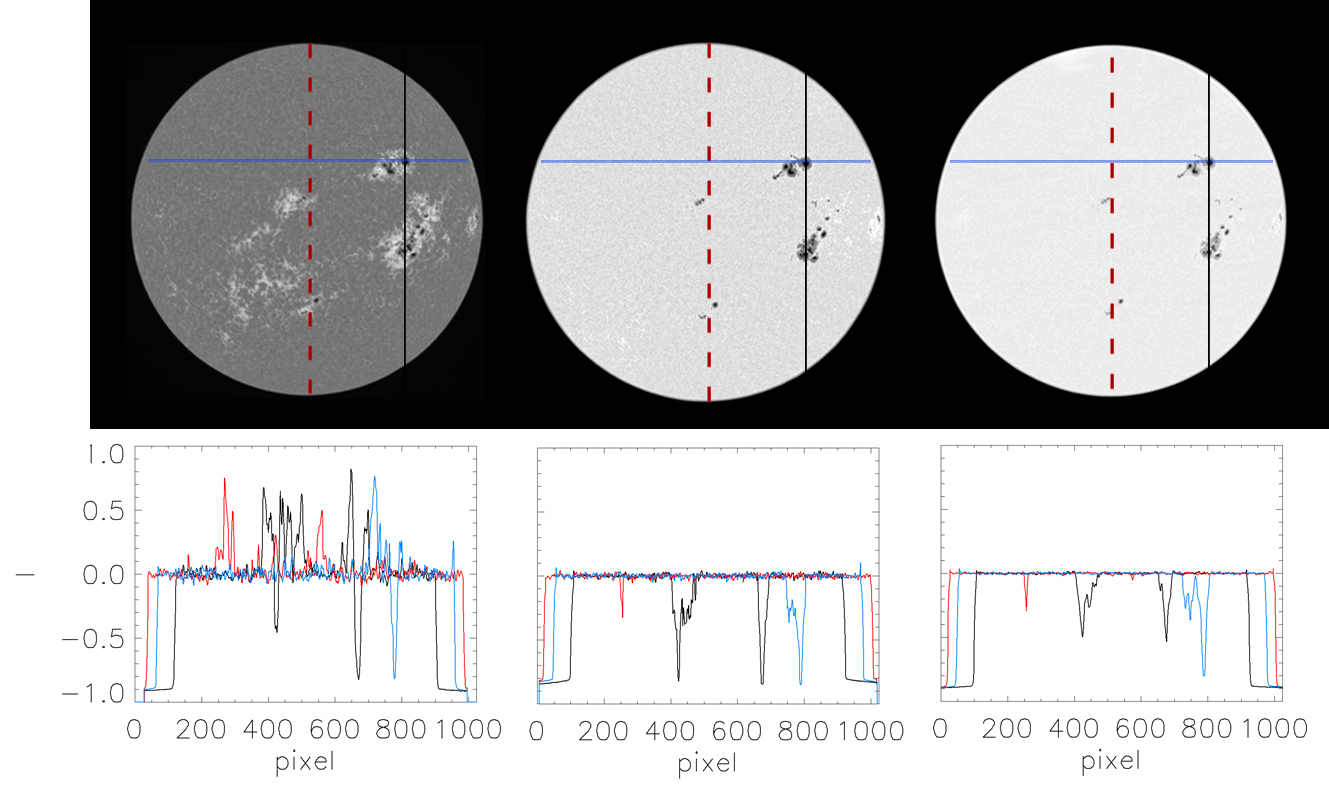}
\end{center}
\caption{
Same as Figure \ref{f6} but for limb-darkening compensated images. 
}\label{f62}
\end{figure*}

\begin{figure*}
\begin{center}
\includegraphics[width=0.95\linewidth]{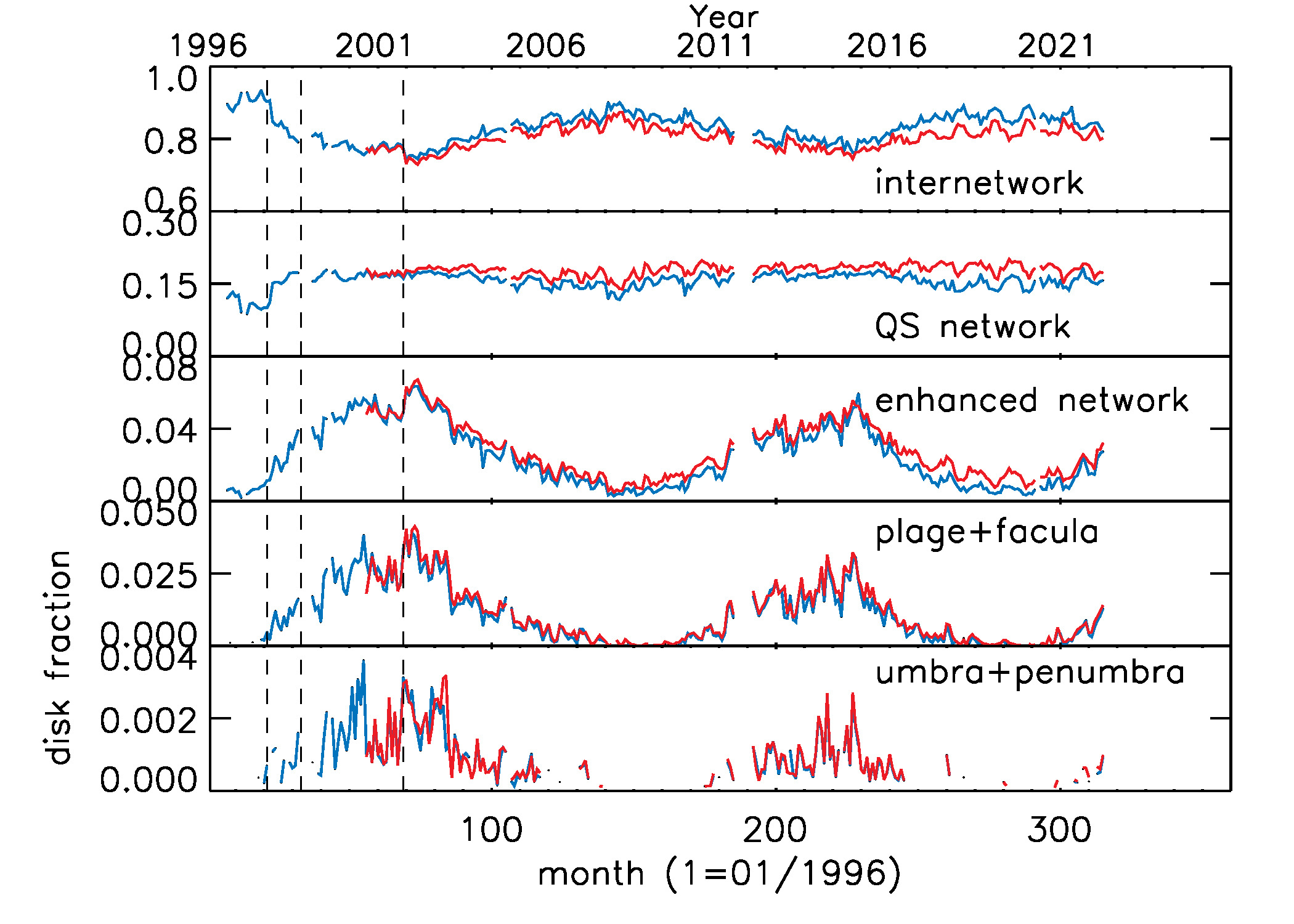}
\end{center}
\caption{Evolution of fractional areas of the seven classes of solar features identified on the Rome/PSPT observations acquired from 1996 to 2022. Lines show monthly median values; red and blue colors are used to show  results from single exposure and sum of 25 exposure observations, respectively. 
}\label{f12}
\end{figure*}

Rome/PSPT observations were aimed to advance our knowledge of solar irradiance changes. However, early data were mostly exploited to  address other research topics.  
In particular, the observations taken over the first few years of operation were employed to investigate 
the geometric, photometric, and oscillation properties of the chromospheric network cells identified in the \ca data  \citep[e.g.][]{ermolli1998,berrilli1998,berrilli1999,pietropaolo1998}. It is worth mentioning that those early studies led to the development of image processing  and skeletonizing methods that  were later applied also to study the properties  of the granulation cells seen in high-resolution observations of the photosphere \citep[e.g.][]{berrilli1999b}, and of the network identified in  Rome/PSPT and other \ca data series \citep[e.g.][]{chatterjee2017}. The Rome/PSPT data were then used to study spatial and temporal variations in
 the photometric properties of the solar disk features \citep[e.g.][]{ermolli2003,ermolli2007}, as a step towards a better understanding of the irradiance changes.  
 However, they have later proved to be valuable also in studying and modelling solar irradiance variations on various time scales.  
Indeed, Rome/PSPT data 
have contributed to show that the measured irradiance changes can be accurately modeled by adding contributions from the magnetic regions emerging on the solar surface in the form of dark and bright features  \citep[e.g.][]{domingo2009,ermolli2011}, and to improve solar irradiance measurements and models   \citep[e.g.][]{ermolli2013}. Moreover, the Rome/PSPT observations were used to reconstruct the rotational and cyclic variations of TSI and SSI with various empirical and semi-empirical models. For example, 
the semi-empirical Solar Radiation Physical Model (SRPM) model \citep[e.g.][]{fontenla1999,fontenla2004,fontenla2006,fontenla2009,fontenla2011,penza2003,penza2004,domingo2009,ermolli2004,ermolli2011,ermolli2013} has largely benefited from use of Rome/PSPT observations in the \ca line and continuum.
Furthermore,  irradiance changes measured since 1996 were accurately reproduced by using 
an empirical model based on the photometric sums (disc integrated intensity after removing limb-darkening and large-scale artefacts) from Rome/PSPT \ca and continuum observations \citep[e.g.][]{chatzistergos2020}. 
The magnetic field strength was also reliably recovered from Rome/PSPT \ca observations \citep[e.g.][]{chatzistergos2019b,Shin2020}, while maps of the magnetic field strength derived from Rome/PSPT \ca observations were used  \citep[e.g.][]{chatzistergos2021} as input to the Spectral and Total irradiance reconstruction \citep[SATIRE, e.g.][]{krivova2003,yeo2014} semi-empirical model. It is worth mentioning that, due to the co-temporality among Rome/PSPT observations, irradiance measurements from space, and continued observations performed with spectroheliographs as in historical solar observations,  
the use of Rome/PSPT images for modelling TSI and SSI variations is an important asset, because it allows setting standards for the analysis of historical data and advance on the modelling of irradiance variations at times prior to existing space-measurements. To this aim, Rome/PSPT data were also widely employed 
to develop accurate methods for processing modern and  historical full-disk observations \citep[e.g.][]{criscuoli2008,ermolli2009,chatzistergos2018}, and to determine the long-term variations in plage areas \citep[e.g.][]{chatzistergos2019p}. Indeed, due to the high quality of Rome/PSPT \ca data, they also acted as the reference data set for the first plage area composite series, which extends back to 1892 \citep{chatzistergos_analysis_2020}.  
Finally, Rome/PSPT observations have also helped with the analysis of TSI and SSI measurements \citep[e.g.][]{harder2019,harder2022}, and of  the impact of solar spectral irradiance variability on Earth's atmosphere \citep[e.g.][]{Merkel2011,ermolli2013}.
In addition, Rome/PSPT observations have    been used as contextual data in the analysis of the small-scale processes revealed by high-resolution observations of the solar atmosphere \citep[e.g.][]{ermolli2022}.
 
Nevertheless, the Rome/PSPT, as its twin MLSO/PSPT were constructed and operated for precision photometric imaging of solar disk features. 
These are evident in the calibrated data and even more so in the images compensated for the solar center-to-limb intensity variation and any eventual large-scale intensity inhomogeneities affecting the data.  
Figure \ref{f62}  shows examples of the Rome/PSPT solar full-disk observations acquired on 31st October 2003 at the \ca line, blue continuum, and red continuum that are shown in Figure \ref{f6},   
after the above compensations. The flat solar images shown in Figure \ref{f62} are usually referred to as \textit{contrast} images. Here, the above compensations, as well as the solar limb determination in the image preprocessing, have been applied with the methods developed by \citet{chatzistergos2018,chatzistergos2020}.   
Contrast values  range [40,80]\%, [30,70]\%, and [5,30]\%  for sunspot, plage, and network regions in the chromospheric \ca observation,  [70,80]\%, [10,30]\%, and [2,4]\% for umbral, penumbral and facular regions in the photospheric blue continuum data, and [50,60]\%, [5,30]\%, and [0.5,1]\% for same regions in the photospheric red continuum images.

We must recall that  
 the combined evolution of  solar disk features such as those described above  reproduce most of the measured irradiance changes. Therefore, an accurate identification and characterization of the various structures seen above the solar disk  is an important step towards a better understanding of the solar irradiance changes.   
To this aim, 
we have  processed the complete series of Rome/PSPT observations with the methods by  \citet{chatzistergos2018,chatzistergos2020}, which resulted to be more precise than other methods in the literature \citep[see, e.g.][]{chatzistergos2022}. 
We then segmented the obtained  \textit{contrast} images  by assuming  seven classes of solar features.  
 These comprise five classes of features identified on the chromospheric \ca observations to localize pixels of average median quiet Sun (internetwork,
QS hereafter), network (QS network), enhanced
network, plage, and bright plage (or facula) regions, and two classes of features identified on the photospheric red continuum data, which correspond  to umbral and penumbral regions. Pixels on the analysed \textit{contrast} images  are identified as a member of one of the above seven classes based on fixed contrast thresholds. 
These are defined for each feature class at 10 heliocentric angles following the atmosphere models and image decomposition method of the SRPM semi-empirical irradiance model by \citet{fontenla1999}.

Figure \ref{f12} shows the evolution of the fractional areas of the seven classes of solar disk features identified on the Rome/PSPT observations acquired from 1996 to 2022 as  single exposure (red) and sum of 25 exposures (blue). Analysis of the two data series produces slightly different results, due to the impact of the slightly different quality of the two data series in these investigations. However, the disk fraction of internetwork regions identified in both data sets show a clear anti-phase variation with solar activity, with disk fraction values in the range $\approx$ [0.73,0.87]. On the other hand, the disk coverage of all the other features identified in the two sets of observations show an in-phase variation with solar activity. This variation is marginally evident for QS network regions identified on sum of 25 exposures observations, while increasingly more pronounced for the other features with monthly median values in the range [0.14,0.20], [0.47,6.7]$\times 10^{-2}$, [0.03,4]$\times 10^{-2}$, and [0.08,3.2]$\times 10^{-3}$ for network, enhanced network, plage and facula, and umbra and penumbra regions, respectively.

The above is an example of  data products derived from Rome/PSPT observations that can enter new studies of the solar irradiance variations over solar cycles 23--25 and longer-term  scales,  which will be presented in future papers.

\section{Conclusions}

The Rome/PSPT program started in 1995 to provide accurate solar full-disk  observations with the main objective of  enhancing our understanding of the solar irradiance variations. Since then 
the Rome/PSPT has generated a valuable time series of photometric full-disk observations of the solar photosphere and chromosphere at various spectral bands. Regular observations started in July 1996 and continued to March 2022, across the solar cycles 23--25.
The observations have been performed daily at the \ca line at 393.3 nm, G-band at 430.6 nm, as well as in red and blue parts of continuum at 409.4 nm and 607.2 nm, respectively. 
For a shorter period (18 months) also observations at the green continuum at 535.7 nm were performed. Over the whole period, the telescope has been operated under three slightly different  set-ups, with a discontinuity in September 1997, and final instrumental characteristics  from September 2001 onward. The Rome/PSPT data archive consists of more than 30000 images  and 1.5 Tb of data. 

The Rome/PSPT observations have proved to be very valuable for studying changes in disk position, area, as well as photometric properties of various solar features. 
Besides, due to the high photometric precision of Rome/PSPT data they have been widely used for reconstructions of TSI and SSI variations, while they also acted as the reference series for the only currently available plage area composite series extending back to 1892. Moreover, Rome/PSPT data  have served to study a variety of other topics, as e.g. the geometric and oscillation properties of the chromospheric network cells, the accuracy of solar irradiance models and measurements, and their impact on the Earth's climate. Furthermore, Rome/PSPT observations have been unique in allowing to bridge series of historical and modern full-disk solar observations, especially as to the \ca images.

The Rome/PSPT is one of the few solar full-disk facilities currently operated and which have observations overlapping the historical series of data made with spectroheliographs since late 19th century, and the  space-based measurements of the solar irradiance. Therefore, Rome/PSPT data represent a unique resource for studies of short- and long-term solar variability, and of solar activity.  
The Rome/PSPT and its observations have helped advance our understanding on solar variability and activity. 
Hopefully, they will continue to do so in the future.

\section*{Conflict of Interest Statement}

The authors declare that the research was conducted in the absence of any commercial or financial relationships that could be construed as a potential conflict of interest.

\section*{Author Contributions}


IE wrote the first draft of the manuscript. FG and TC wrote sections of the manuscript. All authors contributed to manuscript revision, read, and approved the submitted version.

\section*{Funding}
This work was supported by the Italian Ministries of Environment and Research, Italian MIUR-PRIN grants 1998, 2000, 2002, 2004, and grant 2017 ''Circumterrestrial Environment: Impact of Sun--Earth Interaction'', 
by Regione Lazio, and by the
European Union's Horizon 2020 research and Innovation program  under grant agreement No 824135 (SOLARNET) and No 739500 (PRE-EST), and FP7 program under grant agreement No 312495 (SOLARNET).

\section*{Acknowledgments}
The Rome/PSPT prototype was realized thanks to the commitment in the project of Dr Massimo Fofi (INAF-OAR) and Prof Bruno Caccin (Universit\`a degli Studi di Roma ``Tor Vergata'').  
The five Rome/PSPT observers are: Massimo Fofi, Ilaria Ermolli, Mauro Centrone, Cinzia Fazzari, Fabrizio Giorgi. We are grateful to them, in particular to Fabrizio Giorgi who has run observations and telescope maintenance since 2009, and to other colleagues who helped to operate the telescope and analyse the data across the years; special thanks also to Giorgio Viavattene for technical work in the ongoing instrumental upgrade.  
	We thank the International Space Science Institute (ISSI, Bern, CH) for support to work of several teams (id: 64, 119, 335, 420, 475) and  meetings that contributed to improve the  Rome/PSPT monitoring program and data products. 
	TC acknowledges support by the German Federal Ministry of Education and Research (Project No. 01LG1909C).  
	This research has made use of NASA's Astrophysics Data System.



\bibliographystyle{Frontiers-Harvard} 
\bibliography{biblio_pspt}




\end{document}